\documentclass[aps, prb, twocolumn, superscriptaddress, preprintnumbers]{revtex4-2}
\usepackage{bm, amsmath, amsfonts, amssymb, ascmac, mathtools, braket}
\usepackage{multirow}
\usepackage{graphicx}
\usepackage{float, color, xcolor}

\usepackage[whole]{bxcjkjatype} 
\usepackage{subfigure} 
\usepackage{amscd} 

\usepackage{bbm}
\usepackage{tabularx}

\usepackage{comment}

\definecolor{rred}{rgb}{0.8, 0.0, 0.0}
\definecolor{bblue}{rgb}{0.0, 0.0, 0.8}

\usepackage[
pagebackref=false,
colorlinks=true,
linkcolor=bblue,
urlcolor=bblue,
filecolor=black,
citecolor=rred,
pdfstartview=FitV,
pdftitle={},
pdfauthor={},
pdfsubject={},
pdfkeywords={},
pdfpagemode=None,
bookmarksopen=true
]{hyperref}

\newcommand{\ii}{\text{i}}

\newcommand{\momentum}{k}

\begin{document}

\title{Chiral quantum chaos around exponentially many zero modes \\ 
in the quantum breakdown model}

\author{Kohei Kawabata}
\email{kawabata@issp.u-tokyo.ac.jp}
\affiliation{Institute for Solid State Physics, University of Tokyo, Kashiwa, Chiba 277-8581, Japan}

\author{Kinya Guan}
\affiliation{Department of Physics, Graduate School of Science, The University of Tokyo, Tokyo 113-0033, Japan}

\author{Hosho Katsura}
\affiliation{Department of Physics, Graduate School of Science, The University of Tokyo, Tokyo 113-0033, Japan}
\affiliation{Institute for Physics of Intelligence, The University of Tokyo, Tokyo 113-0033, Japan}
\affiliation{Trans-Scale Quantum Science Institute, The University of Tokyo, Tokyo 113-0033, Japan}

\date{\today}

\begin{abstract}
The quantum breakdown model is a model of randomly interacting fermions,
motivated by the physics of dielectric breakdown.
Here, we establish classification of symmetry and quantum chaos in a zero-dimensional, all-to-all-interacting version of the quantum breakdown model. 
It exhibits the $\mathbb{Z}_4$ periodicity with respect to the number of fermionic modes,
reminiscent of the symmetry classification of the Sachdev-Ye-Kitaev model.
A unique feature of the quantum breakdown model is that it realizes all the five classes with chiral symmetry and hosts an exponentially large number of many-body zero modes protected by the chiral index.
We elucidate the separation and scaling of the spectral gap and demonstrate the symmetry-enriched hard-edge spectral statistics as signatures of quantum chaos in the chiral symmetry classes.
\end{abstract}

\maketitle

\section{Introduction}

The spectral statistics provide characterization of quantum chaos and underlie the fundamental understanding of quantum statistical mechanics~\cite{Haake-textbook, Huse-review, Rigol-review}.
Whereas integrable systems exhibit Poisson statistics and lack level repulsion~\cite{Berry-Tabor-77},
the local spectral correlations of chaotic systems are expected to follow those of Hermitian random matrices~\cite{BGS-84}.
Here, the relevant random-matrix universality class is determined by internal symmetry.
In the spectral bulk, the presence or absence of time-reversal symmetry, 
as well as its sign, 
gives rise to the threefold Wigner-Dyson classification~\cite{Wigner-51, *Wigner-58, Dyson-62, Mehta-textbook}. 
Time-reversal symmetry with positive or negative sign leads to the orthogonal or symplectic class, respectively,
whereas its absence leads to the unitary class.
The corresponding level repulsion is characterized by the Dyson index $\beta = 1, 2, 4$.

Chiral and particle-hole symmetries impose additional constraints that become manifest only around the spectral origin~\cite{Gade-91, *Gade-93, Verbaarschot-94, *Verbaarschot-00-review, AZ-97}. 
They reverse the sign of the Hamiltonian through unitary and antiunitary transformations, respectively, 
pairing positive and negative eigenenergies and distinguishing zero energy from a generic point in the spectral bulk. 
The resulting correlations control how energy levels approach the origin,
including the power-law behavior of the density of states and the distribution of the smallest positive eigenenergy. 
The associated exponent $\alpha$ depends on the symmetry class and the chiral index $\nu$ specifying protected zero modes. 
Combining time-reversal, particle-hole, and chiral symmetries defines the tenfold Altland-Zirnbauer classification~\cite{AZ-97}. 
This classification organizes not only Hermitian random matrices but also Anderson transitions~\cite{Beenakker-review-97, *Beenakker-review-15, Evers-review} and topological insulators and superconductors~\cite{Schnyder-08, *Ryu-10, Kitaev-09, HK-review, QZ-review, CTSR-review}. 

The Sachdev-Ye-Kitaev (SYK) model provides a prototypical interacting realization of the connection between symmetry and quantum chaos~\cite{Sachdev-Ye-93, Kitaev-KITP15}. 
Its standard formulation consists of Majorana-fermionic modes coupled through random all-to-all quartic interactions,
whereas its complex-fermion counterpart involves particle-number-conserving interactions~\cite{Sachdev-15}. 
Their connections to strongly correlated matter and black-hole physics, 
as well as their random-matrix spectral correlations, 
have made the SYK-type models central to both condensed-matter and high-energy physics~\cite{Polchinski-Rosenhaus-16, Maldacena-Stanford-16, Gu-17, Fu-17, Song-17, Rosenhaus-review, Sachdev-review}. 
Moreover, the symmetry classification of the SYK-type models fits within the Altland-Zirnbauer framework and depends on the number of fermionic modes and fermion parity~\cite{You-17, Fu-16, GarciaGarcia-16, Cotler-17, Li-17, Kanazawa-17, Behrends-19, Sun-20}.
The Majorana SYK model exhibits the eightfold periodicity in the number of Majorana modes, 
while the complex counterpart exhibits the fourfold periodicity. 
Thanks to these periodic structures,
the SYK-type models represent a useful platform for investigating quantum chaos enriched by symmetry.

Recently, a related model of randomly interacting fermions, 
called the quantum breakdown model, 
was introduced as a quantum description of dielectric breakdown~\cite{Lian-23}. 
Its elementary process converts one fermion into three, 
in addition to the reverse process required by Hermiticity. 
These conversion processes accompany motion between neighboring sites and mimic the amplification of a microscopic excitation into a particle avalanche. 
The breakdown interactions change particle number while preserving fermion parity, 
which is similar to the Majorana SYK model but different from the complex SYK model.
A disorder-free all-to-all version has also been shown to be exactly solvable, 
with its spectral form factor and out-of-time-ordered correlator evaluated analytically~\cite{Guan-Katsura-26}.
While these results demonstrate the rich structure of the quantum breakdown model,
how its internal symmetry organizes the universal spectral statistics has remained largely unexplored.
In particular, the similarities to and differences from the SYK model have remained to be elucidated from the perspective of symmetry-resolved quantum chaos.

In this work, we develop systematic classification of internal symmetry and quantum chaotic spectral statistics of a zero-dimensional version of the quantum breakdown model with random all-to-all interactions. 
This zero-dimensional setting allows us to isolate the distinctive role of the breakdown interactions from additional effects, 
such as the spatial structure and one-body potentials. 
We show that the interplay between fermion parity and antiunitary symmetries gives rise to the fourfold periodicity in the number of complex fermions, 
as summarized in Table~\ref{tab:symmetry}.
The resulting classification thus shares the periodicity of the complex SYK model, 
despite the absence of particle-number conservation. 
A crucial difference from the SYK model is the emergence of chiral symmetry in the many-body Hamiltonian. 
Varying the number of fermionic modes and fermion parity, 
we find that the same breakdown interactions realize all the five Altland-Zirnbauer classes possessing chiral symmetry: 
AIII, BDI, CI, CII, and DIII.

Furthermore, we show that the dimensional imbalance between the two chiral subspaces can grow exponentially with the number of fermionic modes. 
The resulting chiral index enforces an exponentially large number of many-body eigenstates with exact zero energy in every disorder realization. 
Although the zero-mode degeneracy grows exponentially, 
it occupies an exponentially vanishing fraction of the entire Hilbert space. 
We find that such a subextensive scaling relative to the Hilbert-space dimension yields the pronounced separation of energy scales and the unique spectral statistics.
Specifically, through exact diagonalization, we demonstrate that the energy spectrum exhibits universal quantum chaotic correlations both in the bulk and near the origin. 
The bulk level-spacing ratios follow the three Wigner-Dyson distributions, 
while the hard-edge statistics resolve the additional information associated with chiral symmetry and the zero-mode structure. 
In particular, two fermion-parity subspaces can share the same bulk statistics while exhibiting distinct distributions of the smallest positive eigenenergy.
These hard-edge spectral statistics elucidate the symmetry- and index-dependent distributions across all the five classes and connect the exact zero-mode counting to the chaotic many-body energy spectrum.

Additionally, we find that the subspaces with the large chiral indices develop an enhanced spectral gap separating the zero modes from the surrounding nonzero energy levels. 
We establish a hierarchy of spectral scales: 
the gap grows relative to the local mean level spacing as the chiral index increases, 
while becoming smaller relative to the full spectral width. 
For even numbers of fermionic modes, 
the enhanced gap is selective with respect to fermion parity, 
with the subspace exhibiting the larger separation determined by the chiral index. 
Moreover, after rescaling by the effective chiral index and the local mean level spacing, 
the nonzero density of states near the gap collapses onto a common square-root profile, 
consistent with random matrices in the chiral Gaussian ensemble.
The mean density of states thus exhibits a common leading form across the large-index chiral ensembles,
even though the microscopic spectral fluctuations retain their dependence on different symmetry classes. 
These results demonstrate how chiral symmetry can support an exponentially large exact degeneracy while reorganizing the surrounding quantum chaotic spectrum on an energy scale much larger than the local level spacing.

Notably, quantum chaos in the presence of a macroscopic zero-mode sector has recently been investigated in a kinetically constrained spin chain~\cite{Jonay-26}.
Macroscopically degenerate many-body zero modes were also studied in the contexts of quantum many-body scars~\cite{Turner-18, Banerjee-21, Buijsman-22} and supersymmetric SYK-type models~\cite{Fu-17, Kanazawa-17, Sannomiya-17}.
By contrast, their coexistence with quantum chaotic spectral correlations and a symmetry-controlled spectral gap described by chiral random matrix theory constitutes a distinctive feature.
Our work identifies this phenomenon in disordered interacting fermions and relates it to the systematic symmetry classification across all the five classes possessing chiral symmetry.

The remainder of this work is organized as follows. 
In Sec.~\ref{sec:model}, we introduce the quantum breakdown model. 
In Sec.~\ref{sec:symmetry}, we establish its symmetry classification. 
In Sec.~\ref{sec:zero}, we derive the chiral indices and determine the number of many-body zero modes.
In Secs.~\ref{sec:bulk} and \ref{sec:edge}, we investigate the spectral statistics in the bulk and around zero energy, respectively. 
In Sec.~\ref{sec:gap}, we study the separation and scaling of characteristic energy scales.
We conclude in Sec.~\ref{sec:conclusion}.
In Appendix~\ref{appendix:zero}, we investigate a disorder-free counterpart of the quantum breakdown model.

\section{Quantum breakdown model}
    \label{sec:model}

We consider a zero-dimensional version of the quantum breakdown model with random all-to-all interactions among $N$ complex fermionic modes~\cite{Lian-23, Guan-Katsura-26}. 
The fermion annihilation operators $c_i$'s and creation operators $c_i^\dagger$'s ($i=1,\ldots,N$) satisfy the canonical anticommutation relations:
\begin{equation}
    \{ c_i, c_j \} = \{ c_i^{\dag}, c_j^{\dag} \} = 0, \quad \{ c_i, c_j^{\dag} \} = \delta_{ij}.
\end{equation}
The Hamiltonian is
\begin{equation}
H =\frac{1}{N}\sum_{i=1}^{N}
\sum_{\substack{1\le j<k<l\le N\\j,k,l\ne i}} \left(
K_{ijkl}c_i^\dagger c_jc_kc_l
+K_{ijkl}^{*}c_l^\dagger c_k^\dagger c_j^\dagger c_i
\right)
    \label{eq:breakdown_model}
\end{equation}
with the complex-valued random coupling $K_{ijkl} \in \mathbb{C}$.
The term $c_i^\dagger c_jc_kc_l$ converts three fermions into one, 
while its Hermitian conjugate $c_l^\dagger c_k^\dagger c_j^\dagger c_i$ describes the reverse process.
In the original quantum breakdown model~\cite{Lian-23}, 
the conversion of one fermion into three accompanies motion between neighboring sites and describes dielectric breakdown. 
Here, the modes have no spatial arrangement, and the Hamiltonian contains only the conversion processes, without quadratic hopping or potential terms. 
This setting isolates the spectral consequences of the breakdown interactions and the corresponding internal symmetry from the spatial structure.

The fermionic random quartic interactions appear reminiscent of those in the complex SYK model, 
although the latter involve terms of the form $c_i^{\dag} c_j^{\dag} c_k c_l$ and preserve particle number.
Meanwhile, the quantum breakdown interactions $c_i^\dagger c_jc_kc_l$'s are included in the quartic interaction terms of the Majorana SYK model.
The ordering $j<k<l$ for $K_{ijkl}$ prevents counting of permutations of the three annihilation indices. 
We additionally impose $j,k,l\ne i$, 
so that each interaction involves four distinct fermionic modes. 
In this respect, the present model differs from the exactly solvable disorder-free all-to-all model~\cite{Guan-Katsura-26}. 
This difference is important for symmetry and zero modes, as discussed below (see also Appendix~\ref{appendix:zero} for details).
Moreover, choosing the complex coupling avoids imposing additional symmetry under complex conjugation that would arise for the real coupling.

For each allowed set of indices, we independently draw the complex coupling $K_{ijkl}$ according to
\begin{equation}
K_{ijkl}
=J\sqrt{\frac{3}{4N}}
\left(x_{ijkl}+\ii y_{ijkl}\right),
    \label{eq:random_couplings}
\end{equation}
with the interaction scale $J > 0$. 
Here, $x_{ijkl}$'s and $y_{ijkl}$'s are mutually independent real Gaussian variables with zero mean and unit variance.
The resulting couplings satisfy
\begin{equation}
\braket{K_{ijkl}}=\braket{K_{ijkl}^{\,2}}=0,
\qquad
\braket{|K_{ijkl}|^2}=\frac{3J^2}{2N},
    \label{eq:coupling_moments}
\end{equation}
where the bracket $\braket{}$ denotes the disorder average.
In addition to the prefactor $1/N$ in Eq.~\eqref{eq:breakdown_model}, 
this choice gives a variance $3J^2/2N^3$ for each four-fermion interaction coefficient, 
ensuring a well-defined large-$N$ limit.
The presence of the random coupling constitutes another clear difference from the disorder-free case~\cite{Guan-Katsura-26}.
Notably, the randomness is introduced through the four-fermion couplings, 
rather than by independently sampling all many-body matrix elements as in conventional random-matrix ensembles. 
Hence, the Gaussian distribution of the couplings does not by itself imply random-matrix spectral statistics.

Each interaction changes the total fermion number by two, 
and thus particle-number conservation is broken. 
Still, fermion parity remains preserved:
\begin{equation}
    [H,(-1)^{\mathcal F}]=0, \quad 
    (-1)^{\mathcal F}
    \coloneqq \prod_{i=1}^{N}\left(1-2c_i^\dagger c_i\right).
    \label{eq:fermion_parity}
\end{equation}
Consequently, the many-body Hilbert space  decomposes into the two subspaces with even and odd fermion parity, 
labeled by $(-1)^{\mathcal F}=+1$ and $(-1)^{\mathcal F}=-1$, respectively, each of dimension $2^{N-1}$. 
We analyze the spectrum separately within individual subspaces of fermion parity. 
We investigate additional unitary and antiunitary symmetries, and their dependence on $N$ and fermion parity in Sec.~\ref{sec:symmetry}.

\begin{figure}[t]
\centering
\includegraphics[width=1.0\linewidth]{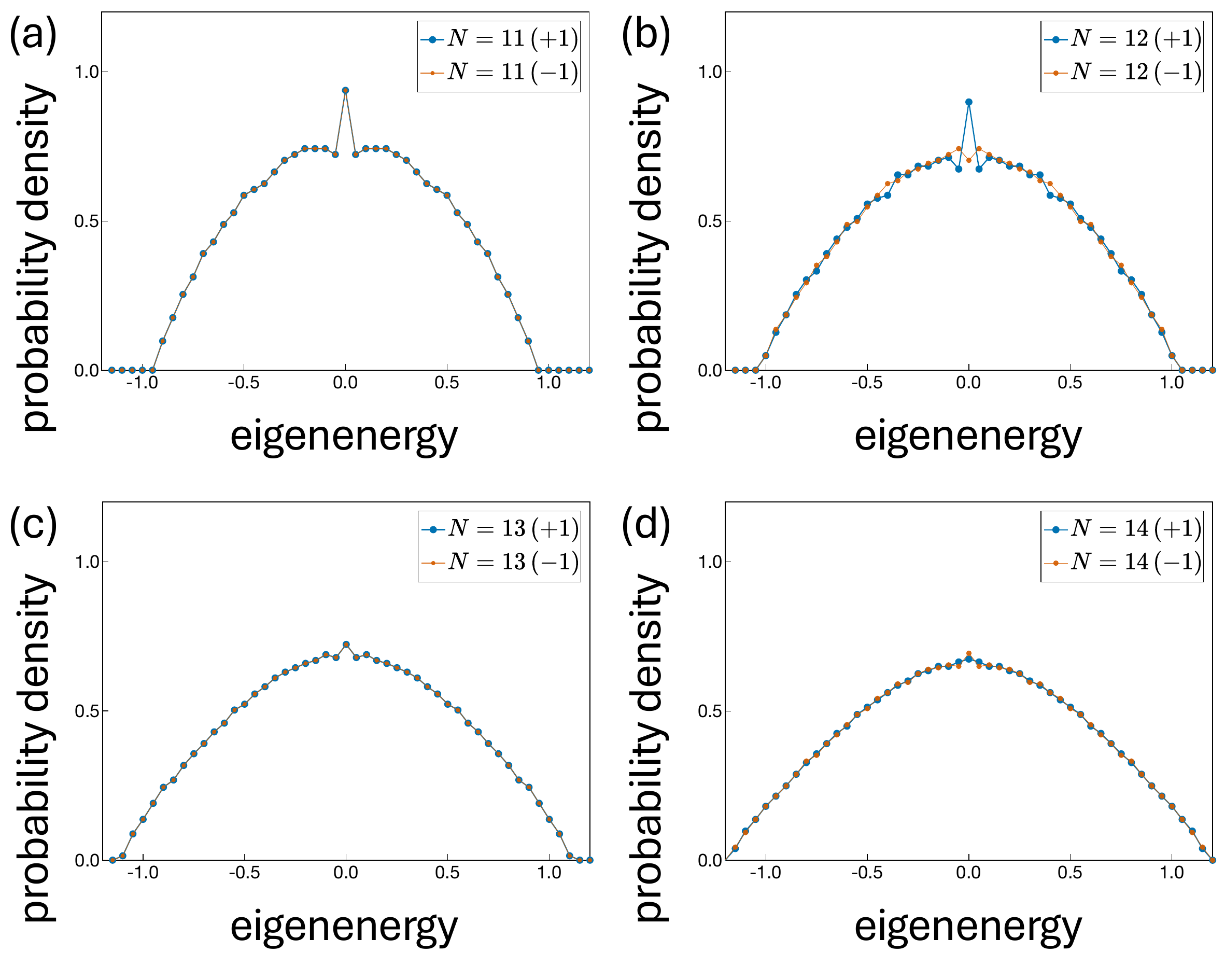} 
\caption{Density of states for a single realization of the quantum breakdown model ($J=1$) with (a)~$N=11$, (b)~$N=12$, (c)~$N=13$, and (d)~$N=14$.
Blue dots: $\left( -1 \right)^{
\cal F} = +1$.
Orange dots: $\left( -1 \right)^{
\cal F} = -1$.}	
    \label{fig:DOS}
\end{figure}

In Fig.~\ref{fig:DOS}, we show the density of states obtained by exact diagonalization for a single disorder realization at each of $N=11,12,13,14$, 
with the two fermion-parity subspaces shown separately. 
The densities are normalized to unit integral within each subspace. 
The nonzero spectrum forms a broad distribution on both sides of zero energy,
and the spectrum within each fermion-parity subspace is symmetric under $E\mapsto-E$. 
This spectral reflection property follows from chiral symmetry, as discussed below. 
For odd $N$, the even and odd fermion-parity subspaces are exactly isospectral, 
and their densities coincide in Figs.~\ref{fig:DOS}\,(a) and (c). 
For even $N$, on the other hand, their broad spectral envelopes remain similar, 
although the two spectra are generally different.

A distinctive feature arises near zero energy. 
Although the broad bulk spectrum is qualitatively similar for all the cases, 
the spectral features around zero energy depend sensitively on $N$ and $\left( -1 \right)^{\cal F}$.
The central peaks, particularly visible in both $\left( -1 \right)^{\cal F} = \pm 1$ subspaces for $N=11$ and in the $\left( -1 \right)^{\cal F} = +1$ subspace for $N=12$, reflect exact many-body zero modes. 
It should be noted that a less prominent zero-energy peak for larger $N$ does not necessarily imply fewer zero modes, 
since the normalized density measures their fraction of the Hilbert-space dimension rather than their absolute number. 
As shown below, this fraction can decrease even when the number of zero modes grows exponentially with $N$. 
The overall density alone, however, neither establishes quantum chaos nor resolves the characteristic separation between the zero modes and the surrounding nonzero levels in the spectral bulk. 
These observations motivate a more detailed analysis of the spectral statistics both in the bulk and near zero energy, 
as well as their symmetry classification.

\section{Symmetry classification}
    \label{sec:symmetry}

\begin{table}[t]
	\centering
	\caption{Periodic table of the quantum breakdown model for the number $N$ (mod $4$) of complex fermions. 
    The entries for the antiunitary symmetries ${\cal P}$ and ${\cal R}$ specify the signs $\pm 1$ of ${\cal P}^2$ and ${\cal R}^2$, respectively.}
     \begin{tabular}{c|cccc} \hline \hline
     ~~$N$ (mod $4$)~~ & ~~$0$~~ & ~~$1$~~ & ~~$2$~~ & ~~$3$~~  \\ \hline
     ${\cal P}$ & $+1$ & $+1$ & $-1$ & $-1$ \\
     ${\cal R}$ & $+1$ & $-1$ & $-1$ & $+1$ \\ \hline 
     ~~~$\left( -1 \right)^{\cal F} = +1$~~~ & ~~~BDI~~~ & \multirow{2}{*}{~~~AIII~~~} & ~~~DIII~~~ & \multirow{2}{*}{~~~AIII~~~} \\
     $\left( -1 \right)^{\cal F} = -1$ & CI & & CII & \\ \hline \hline
    \end{tabular}
	\label{tab:symmetry}
\end{table}

\begin{table}[t]
	\centering
	\caption{Tenfold Altland-Zirnbauer symmetry classification based on time-reversal, particle-hole, and chiral symmetries.
    For the antiunitary symmetries of time reversal and particle-hole transformation, 
    whereas ``$0$" denotes their absence,
    ``$+1$" and ``$-1$" denote their presence with the corresponding signs.
    For the unitary symmetry of chiral transformation,
    ``$0$" and ``$1$" denote its absence and presence, respectively.}
	\label{tab:AZ}
     \begin{tabular}{cccc} \hline \hline
    ~~Class~~ & ~~Time reversal~~ & ~~Particle hole~~ & ~~Chiral~~ \\ \hline
    A & $0$ & $0$ & $0$ \\
    AIII & $0$ & $0$ & $1$ \\ \hline
    AI & $+1$ & $0$ & $0$ \\
    BDI & $+1$ & $+1$ & $1$ \\
    D & $0$ & $+1$ & $0$ \\
    DIII & $-1$ & $+1$ & $1$ \\
    AII & $-1$ & $0$ & $0$ \\
    CII & $-1$ & $-1$ & $1$ \\
    C & $0$ & $-1$ & $0$ \\
    CI & $+1$ & $-1$ & $1$ \\ \hline \hline
  \end{tabular}
\end{table}

We now classify symmetry of the quantum breakdown model.  
The interplay between fermion parity, antiunitary symmetry, and chiral symmetry gives rise to the fourfold periodicity in $N$, summarized in Table~\ref{tab:symmetry}. 
See also Table~\ref{tab:AZ} for a summary of the tenfold Altland-Zirnbauer symmetry classification.
Notably, the same form of the breakdown interactions realizes all the five symmetry classes with chiral symmetry,
providing a common setting for studying their distinct spectral correlations near zero energy.

\subsection{Antiunitary symmetry}
    \label{subsec:antiunitary}

We introduce the antiunitary operators
\begin{align}
\mathcal{P}
&\coloneqq\left[\prod_{i=1}^{N}\left(c_i+c_i^\dagger\right)\right]\mathcal{K}, 
    \label{eq:P} \\
\mathcal{R}
&\coloneqq\left[\prod_{i=1}^{N}\ii\left(c_i-c_i^\dagger\right)\right]\mathcal{K},
    \label{eq:R}
\end{align}
where $\mathcal{K}$ denotes complex conjugation.
Counting the fermionic signs in the ordered products and the factors of the imaginary unit $\ii$ under complex conjugation, we obtain
\begin{align}
\mathcal{P}^{2} &= (-1)^{N(N-1)/2} = \begin{cases}
    +1 & \left[ N \equiv 0, 1 \left( \text{mod}~4\right) \right],  \\
    -1 & \left[ N \equiv 2, 3 \left( \text{mod}~4\right) \right];
\end{cases} 
    \label{eq:sign-P} \\
\mathcal{R}^{2} &= (-1)^{N(N+1)/2} =
\begin{cases}
    +1 & \left[ N \equiv 0, 3 \left( \text{mod}~4\right) \right], \\
    -1 & \left[ N \equiv 1, 2 \left( \text{mod}~4\right) \right].
\end{cases}
    \label{eq:sign-R}
\end{align}

These antiunitary operators act on the fermion operators as
\begin{equation}
\mathcal{P}c_i\mathcal{P}^{-1}=(-1)^{N+1}c_i^\dagger,
\qquad
\mathcal{R}c_i\mathcal{R}^{-1}=(-1)^N c_i^\dagger.
\end{equation}
Under either operation, each interaction term is mapped to its Hermitian conjugate,
while the coupling $K_{ijkl}$ is complex conjugated.
Consequently, the Hamiltonian $H$ is invariant under both $\mathcal{P}$ and $\mathcal{R}$:
\begin{equation}
\mathcal{P}H\mathcal{P}^{-1}=H,
\qquad
\mathcal{R}H\mathcal{R}^{-1}=H.
    \label{eq:antiunitary_hamiltonian}
\end{equation}
These relations hold for each individual disorder realization, 
rather than only after disorder averaging.
It should be noted that the constraint that all four indices of $K_{ijkl}$ are distinct allows the fermion operators to be reordered without generating additional contraction terms.
Thus, $\mathcal{P}$ and $\mathcal{R}$  preserve the quantum breakdown model and effectively act as time-reversal symmetry,
with their signs determined by Eqs.~\eqref{eq:sign-P} and \eqref{eq:sign-R}. 

\subsection{Fermion parity}
    \label{subsec:fermion_parity}

We need to study symmetry and spectral statistics within each subspace of fermion parity $\left( -1 \right)^{\cal F}$ in Eq.~\eqref{eq:fermion_parity}.
Otherwise, statistically independent spectra from different symmetry sectors would be mixed,
obscuring the signatures of quantum chaos.
The two antiunitary operations $\mathcal{P}$ and $\mathcal{R}$ are related to fermion parity $\left( -1 \right)^{\cal F}$ through
\begin{equation}
\mathcal{P}\mathcal{R}
=\ii^N(-1)^{N(N-1)/2}(-1)^{\mathcal F},
    \label{eq:antiunitary_product}
\end{equation}
and
\begin{align}
(-1)^{\mathcal F}\mathcal{P}(-1)^{\mathcal F}
&=(-1)^N\mathcal{P},\\
(-1)^{\mathcal F}\mathcal{R}(-1)^{\mathcal F}
&=(-1)^N\mathcal{R}.
\end{align}

Thus, for even $N$, both antiunitary operations preserve each subspace of $\left( -1 \right)^{\cal F}$ and remain symmetry of the corresponding block of the Hamiltonian.
Equation~\eqref{eq:antiunitary_product} further shows that, 
within this subspace of fixed $\left( -1 \right)^{\cal F}$, 
the two antiunitary symmetries differ only by a phase and do not impose independent constraints.
For odd $N$, on the other hand, both operations exchange the two subspaces of $\left( -1 \right)^{\cal F} = +1$ and $\left( -1 \right)^{\cal F} = -1$. 
Therefore, $\mathcal{P}$ and $\mathcal{R}$ relate the spectra of the two different blocks rather than constrain individual blocks separately. 
Accordingly, they do not directly influence the spectral statistics within the fixed fermion-parity subspace.
In addition to Eq.~\eqref{eq:antiunitary_hamiltonian},
this exchange implies that the two fermion-parity subspaces are exactly isospectral for each disorder realization.

\subsection{Chiral symmetry}
    \label{subsec:chiral}

The quantum breakdown interactions change fermion number only by two.
Accordingly, within the subspace of fermion parity $\left( -1 \right)^{\cal F} = +1$,
they connect many-body fermionic states with $n \equiv 0$ (mod $4$) only to those with $n \equiv 2$;
within the subspace of fermion parity $\left( -1 \right)^{\cal F} = -1$,
they connect many-body fermionic states with $n \equiv 1$ (mod $4$) only to those with $n \equiv 3$.
Here, $n$ is the number of fermions.
This bipartite structure of the many-body Hilbert space gives rise to chiral symmetry in the quantum breakdown model.

Specifically, we introduce a unitary operator
\begin{equation}
\Gamma\coloneqq\exp\left(-\frac{\ii\pi}{2}\sum_{i=1}^{N}c_i^\dagger c_i\right),
\qquad
\Gamma^2=(-1)^{\mathcal F}.
    \label{eq:chiral_operator}
\end{equation}
Its action on the fermion operators is
\begin{equation}
\Gamma c_i\Gamma^{-1}=\ii c_i,
\qquad
\Gamma c_i^\dagger\Gamma^{-1}=-\ii c_i^\dagger,
\label{eq:chiral_fermions}
\end{equation}
yielding chiral symmetry
\begin{equation}
\Gamma H\Gamma^{-1}=-H.
\label{eq:chiral_hamiltonian}
\end{equation}
Chiral symmetry pairs nonzero energy $E$ and $-E$, and changes spectral correlations around zero energy $E=0$.

Since $\Gamma$ commutes with fermion parity, chiral symmetry remains relevant even within each subspace of fermion parity.
In the full Hilbert space, we have $\Gamma^4=1$, while $\Gamma^2$ equals fermion parity $\left( - 1\right)^{\cal F}$ rather than the identity. 
Within the even-parity subspace $\left( - 1\right)^{\cal F} = +1$, 
$\Gamma$ itself is a Hermitian chiral operator satisfying $\Gamma^2 = +1$. 
Within the odd-parity subspace, the corresponding Hermitian chiral operator is $\ii\Gamma$, which also squares to $+1$. 
The additional phase factor $\ii$ is important for determining its algebra with antiunitary symmetry,
as described below.

Applying Eq.~\eqref{eq:chiral_fermions} to Eqs.~\eqref{eq:P} and \eqref{eq:R}, we obtain
\begin{equation}
\Gamma\mathcal{P}\Gamma^{-1}=(-\ii)^N\mathcal{P}, \quad
\Gamma\mathcal{R}\Gamma^{-1}=(-\ii)^N\mathcal{R}.
    \label{eq:chiral_antiunitary}
\end{equation}
Replacing $\Gamma$ by $\ii\Gamma$ gives
\begin{equation}
(\ii\Gamma)\mathcal{P}(\ii\Gamma)^{-1}=-(-\ii)^N\mathcal{P}, \quad
(\ii\Gamma)\mathcal{R}(\ii\Gamma)^{-1}=-(-\ii)^N\mathcal{R},
\label{eq:odd_chiral_antiunitary}
\end{equation}
where the additional minus sign arises from antiunitarity of $\mathcal{P}$ and $\mathcal{R}$. 
For even $N$, this additional sign reverses the commutation relation between the chiral and antiunitary operations upon changing the fermion-parity subspace.
Consequently, the two fermion-parity blocks can belong to different symmetry classes even when the corresponding antiunitary symmetry has the same sign.

\subsection{Symmetry classes}

The interplay of the symmetries discussed above---antiunitary symmetries in Sec.~\ref{subsec:antiunitary}, fermion parity symmetry in Sec.~\ref{subsec:fermion_parity}, and chiral symmetry in Sec.~\ref{subsec:chiral}---determines the relevant symmetry classes of the quantum breakdown model.
Specifically, we have the following symmetry classification according to $N$ (mod $4$):

\begin{itemize}
\item For $N\equiv0\pmod4$, the antiunitary symmetry $\mathcal{P}$ is respected within each subspace of fermion parity $\left( -1 \right)^{\cal F}$ and satisfies $\mathcal{P}^2 = +1$.
In the even-parity subspace $\left( -1 \right)^{\cal F} = +1$, 
$\Gamma$ commutes with $\mathcal{P}$, 
leading to class BDI. 
In the odd-parity subspace $\left( -1 \right)^{\cal F} = -1$, 
on the other hand,
$\ii\Gamma$ anticommutes with $\mathcal{P}$, 
leading to class CI.

\item For $N\equiv1,3\pmod4$, $\mathcal{P}$ and $\mathcal{R}$ merely exchange the two subspaces of fermion parity $\left( -1 \right)^{\cal F}$,
and no longer act as symmetry within each subspace.
Thus, only chiral symmetry $\Gamma$ is relevant,
leading to class AIII for both $\left( -1 \right)^{\cal F} = \pm 1$.

\item For $N\equiv2\pmod4$, the antiunitary symmetry $\mathcal{P}$ is respected within each subspace of fermion parity $\left( -1 \right)^{\cal F}$ and satisfies $\mathcal{P}^2 = -1$.
In the even-parity subspace $\left( -1 \right)^{\cal F} = +1$, 
$\Gamma$ anticommutes with $\mathcal{P}$, 
leading to class DIII. 
In the odd-parity subspace $\left( -1 \right)^{\cal F} = -1$, 
on the other hand,
$\ii\Gamma$ commutes with $\mathcal{P}$, 
leading to class CII.
\end{itemize}

These results establish the periodic table of the quantum breakdown model (Table~\ref{tab:symmetry}).
The symmetry classes repeat with period four in the number $N$ of complex fermionic modes.
Whereas the Majorana SYK model exhibits the $\mathbb{Z}_8$ classification, 
the particle-number-conserving complex SYK model follows the $\mathbb{Z}_4$ classification~\cite{You-17}.
We here show that the quantum breakdown model conforms to the $\mathbb{Z}_4$ classification.
Furthermore, a unique feature of the quantum breakdown model is the emergence of chiral symmetry.
Such a chiral structure also underlies the emergence of many-body zero modes, 
as discussed in Sec.~\ref{sec:zero}. 
This classification also determines the universal spectral correlations both in the bulk (Sec.~\ref{sec:bulk}) and in the vicinity of zero energy (Sec.~\ref{sec:edge}). 
Notably, for even $N$, the two fermion-parity blocks can share the same bulk spectral statistics despite belonging to different symmetry classes, 
whose distinction becomes manifest only around zero energy.

\section{Many-body zero modes}
    \label{sec:zero}

We investigate the many-body eigenstates exactly at zero energy protected by chiral symmetry. 
We show the numerically obtained number of many-body zero modes in Table~\ref{tab:zero},
revealing a strong dependence on both $N$ and fermion parity $\left( -1 \right)^{\cal F}$, 
including degeneracies that grow exponentially with respect to $N$. 
We elucidate these results in terms of two distinct constraints: 
the dimensional imbalance between the chiral subspaces and the additional rank deficiency arising from isolated states of the breakdown interactions. 
Combining these two constraints, 
we analytically derive the lower bounds on the number of many-body zero modes summarized in Table~\ref{tab:zero-v2},
which are saturated for all the cases with $N \leq 16$ in Table~\ref{tab:zero}.
In the following, we count individual many-body eigenstates, 
including both eigenstates of each Kramers pair.
Similar zero-mode counting was also discussed, for example, for quantum spin models~\cite{Schecter-18} and SYK-type models~\cite{Iyoda-18}.

\begin{table}[t]
	\centering
	\caption{Number of many-body zero modes in the quantum breakdown model obtained from the numerical diagonalization.}
     \begin{tabular}{c|cc} \hline \hline
     ~~$N$~~ & ~~$\left( -1 \right)^{\cal F} = +1$~~ & ~~$\left( -1 \right)^{\cal F} = -1$~~  \\ \hline
     $4$ & $8$ & $0$ \\ 
     $5$ & $6$ & $6$ \\ 
     $6$ & $4$ & $8$ \\ 
     $7$ & $8$ & $8$ \\     
     $8$ & $16$ & $0$ \\ 
     $9$ & $16$ & $16$ \\ 
     $10$ & $4$ & $32$ \\ 
     $11$ & $34$ & $34$ \\ 
     $12$ & $68$ & $0$ \\ 
     $13$ & $66$ & $66$ \\ 
     $14$ & $4$ & $128$ \\  
     $15$ & $128$ & $128$ \\  
     $16$ & $256$ & $0$ \\ \hline \hline
    \end{tabular}
	\label{tab:zero}
\end{table}

\begin{table}[t]
	\centering
	\caption{Lower bounds on the number of many-body zero modes in the quantum breakdown model.}
     \begin{tabular}{c|cc} \hline \hline
     ~$N$ (mod $8$)~ & ~~$\left( -1 \right)^{\cal F} = +1$~~ & ~~$\left( -1 \right)^{\cal F} = -1$~~  \\ \hline
     $0$ & $2^{N/2}$ (BDI) & $0$ (CI) \\ 
     $1$ & ~~$2^{(N-1)/2}$ (AIII)~~ & $2^{(N-1)/2}$ (AIII) \\
     $2$ & $4$ (DIII) & ~~$2^{N/2}$ (CII)~~ \\
     $3$ & ~~$2^{(N-1)/2} + 2$ (AIII)~~ & ~~$2^{(N-1)/2} + 2$ (AIII)~~ \\    
     $4$ & $2^{N/2} + 4$ (BDI) & $0$ (CI) \\
     $5$ & ~~$2^{(N-1)/2} + 2$ (AIII)~~ & ~~$2^{(N-1)/2} + 2$ (AIII)~~ \\ 
     $6$ & $4$ (DIII) & $2^{N/2}$ (CII) \\ 
     $7$ & $2^{(N-1)/2}$ (AIII) & $2^{(N-1)/2}$ (AIII) \\ \hline \hline
    \end{tabular}
	\label{tab:zero-v2}
\end{table}

\subsection{Chiral index}

As discussed in Sec.~\ref{sec:symmetry},
the quantum breakdown interactions change the particle number only by two,
giving rise to the chiral structure in the many-body Hilbert space.
In general, in the presence of chiral symmetry, the Hamiltonian $H$ can be expressed in block-off-diagonal form,
\begin{equation}
H=
\begin{pmatrix}
0&h\\
h^\dagger&0
\end{pmatrix},
\label{eq:zero_chiral_block}
\end{equation}
where $h$ is a $p \times q$ matrix.
In the $\left( -1 \right)^{\cal F} = +1$ ($-1$) subspace of the quantum breakdown model, 
$p$ denotes the number of many-body fermionic states with $n \equiv 0$ ($1$) (mod $4$),
and $q$ denotes that with $n \equiv 2$ ($3$) (mod $4$),
where $n$ is the number of fermions.
The positive eigenenergies of $H$ are given as the singular values of $h$.
Accordingly, the number of zero modes is generally given as
\begin{align}
\mathcal{D}_{\mathrm{zero}}
&\coloneqq\dim\ker H \nonumber \\
&=p+q-2\operatorname{rank}h \nonumber \\
&= \left| \nu \right| + 2\bigl[\min(p,q)-\operatorname{rank}h\bigr]
    \label{eq:zero_rank_count}
\end{align}
with the chiral index $\nu \coloneqq p-q$.

A unique feature of the quantum breakdown model is that the chiral index $\nu$ can grow exponentially with $N$.
Specifically, the chiral index $\nu_+$ in the $\left( -1 \right)^{\cal F} = +1$ subspace is given as
\begin{align}
\nu_+\coloneqq{}&
\sum_{\substack{0\le n\le N\\n\equiv0~\left( \text{mod}\,4 \right)}}\binom{N}{n}
-
\sum_{\substack{0\le n\le N\\n\equiv2~\left( \text{mod}\,4 \right)}}\binom{N}{n}
\nonumber \\
={}&2^{N/2}\cos\frac{\pi N}{4},
    \label{eq:zero_chiral_index_even}
\end{align}
and the chiral index $\nu_-$ in the $\left( -1 \right)^{\cal F} = -1$ subspace is given as
\begin{align}
\nu_-\coloneqq{}&
\sum_{\substack{0\le n\le N\\n\equiv1~\left( \text{mod}\,4 \right)}}\binom{N}{n}
-
\sum_{\substack{0\le n\le N\\n\equiv3~\left( \text{mod}\,4 \right)}}\binom{N}{n}
\nonumber \\
={}&2^{N/2}\sin\frac{\pi N}{4}.
    \label{eq:zero_chiral_index_odd}
\end{align}
These closed-form expressions follow from the real and imaginary parts of the identity
\begin{equation}
\nu_++\ii\nu_-=(1+\ii)^N.
\label{eq:zero_index_identity}
\end{equation}
For odd $N$, we have $|\nu_+|=|\nu_-|=2^{(N-1)/2}$.
For $N\equiv0\pmod4$, the chiral index is nonzero only for $\left( -1 \right)^{\cal F} = +1$,
with $|\nu_+|=2^{N/2}$; 
for $N\equiv2\pmod4$, it is nonzero only for $\left( -1 \right)^{\cal F} = -1$,
with $|\nu_-|=2^{N/2}$. 
Thus, the subspace with an exponentially large chiral index switches depending on $N$,
which further enforces an exponentially large number of zero modes.

\subsection{Additional zero modes from rank deficiency}

The structure of the quantum breakdown model can further induce the rank deficiency $\operatorname{rank} h < \min \left( p, q \right)$ in Eq.~\eqref{eq:zero_rank_count}.
Specifically, the fermionic vacuum $\lvert\mathrm{vac}\rangle$ and the fully occupied state $\lvert\mathrm{full}\rangle$ satisfy
\begin{equation}
H\lvert\mathrm{vac}\rangle=H\lvert\mathrm{full}\rangle=0,
    \label{eq:zero_isolated_states}
\end{equation}
and constitute many-body zero modes.
Notably, both $\ket{\rm vac}$ and $\ket{\rm full}$ are completely disconnected from all other occupation-number states.
If such an isolated state belongs to the smaller-dimensional space of particle number $n$ (mod $4$),
it removes a row or column that would otherwise contribute to the maximal rank of $h$.
Each such independent rank deficiency 
[i.e., $\min \left( p, q \right) - \operatorname{rank} h = 1$]
increases $\mathcal{D}_{\rm zero}$ by two according to Eq.~\eqref{eq:zero_rank_count}.
By contrast, if the isolated state belongs to the larger-dimensional space, 
its contribution is already accounted for by the dimensional imbalance,
and no additional rank deficiency is enforced by this isolated state.

Therefore, the correction to the zero-mode count beyond the chiral index is determined at least by whether $\ket{\rm vac}$ and $\ket{\rm full}$ belong to the smaller-dimensional space.
This depends on $N$ (mod $8$), 
as described below.
Note that $\ket{\rm vac}$ always has even fermion parity $\left( -1 \right)^{\cal F} = +1$,
whereas $\ket{\rm full}$ has $\left( -1 \right)^{\cal F} = \left( -1 \right)^N$.
It should also be noted that this counting argument only provides lower bounds on the number of many-body zero modes;
additional mechanisms can yield further zero modes.

\begin{itemize}
\item For $N\equiv0,1,7\pmod8$, both isolated states belong to the larger-dimensional space. 
Thus, their presence does not give any additional contributions beyond those imposed by the chiral index,
leading to the zero-mode counts summarized in Table~\ref{tab:zero-v2}.

\item For $N\equiv2,6\pmod8$, 
in the $\left( -1 \right)^{\cal F} = +1$ subspace (class DIII), 
the two spaces have equal dimensions $p=q=2^{N-2}$,
where $\ket{\rm vac}$ belongs to the space of $n \equiv 0$ (mod $4$) and $\ket{\rm full}$ to that of $n \equiv 2$.
Given the Kramers degeneracy, the rank deficiency is at least $\min \left( p, q \right) - \operatorname{rank} h = 2$, 
yielding $\mathcal{D}_{\rm zero} \geq 4$ (see Table~\ref{tab:zero-v2}).
In the $\left( -1 \right)^{\cal F} = -1$ subspace (class CII),
no isolated states contribute an additional rank deficiency, 
and $\mathcal{D}_{\rm zero}$ is entirely determined by the chiral index $\nu_-=\pm 2^{N/2}$.

\item For $N\equiv3,5\pmod8$, $\ket{\rm vac}$ belongs to the smaller-dimensional space within the $\left( -1 \right)^{\cal F} = +1$ subspace, 
and $\ket{\rm full}$ belongs to the smaller-dimensional space within the $\left( -1 \right)^{\cal F} = -1$ subspace. 
Consequently, each fermion-parity block has a rank deficiency of at least one and contains at least $2^{(N-1)/2}+2$ zero modes.  
For example, the counts $34$ for $N=11$ and $66$ for $N=13$ in Table~\ref{tab:zero} agree with this expression.

\item For $N\equiv4\pmod8$, both isolated states $\ket{\rm vac}$ and $\ket{\rm full}$ belong to the $\left( -1 \right)^{\cal F} = +1$ subspace. 
Specifically, both of them lie in the smaller-dimensional subspace of $n \equiv 0$ (mod $4$) and produce the rank deficiency $\min \left( p, q \right) - \operatorname{rank} h = 2$, 
yielding $\mathcal{D}_{\rm zero} \geq 2^{N/2} + 4$ (see Table~\ref{tab:zero-v2}).
On the other hand, the $\left( -1 \right)^{\cal F} = -1$ subspace has $\nu_-=0$ and contains neither isolated state, 
and hence no zero modes are enforced by these constraints.
\end{itemize}

\subsection{Exponential degeneracy of zero modes}

Combining the chiral-index counting with the isolated-state constraints gives the lower bounds on the number of zero modes,
as summarized in Table~\ref{tab:zero-v2}. 
The numerical zero-mode counts in Table~\ref{tab:zero} saturate these bounds for all $N=4,\ldots,16$. 
Although the symmetry classes repeat with period four in $N$, 
the corrections to the index bounds exhibit the eightfold periodicity. 
This distinction originates from the signs of the chiral indices. 
Increasing $N$ by four can exchange the larger and smaller spaces without changing the symmetry class, 
thereby changing whether the isolated states generate an additional rank deficiency.

For $\nu_\pm\ne0$, 
chiral symmetry alone guarantees an exponentially large number of zero modes.
Specifically, we have
\begin{equation}
\mathcal{D}_{\mathrm{zero}}= \mathcal{O}\,( 2^{N/2} ),
\qquad
\frac{\mathcal{D}_{\mathrm{zero}}}{2^{N-1}} = \mathcal{O}\,( 2^{-N/2} )
\label{eq:zero_exponential_scaling}
\end{equation}
for $N \to \infty$.
While the zero-mode subspace grows exponentially with $N$,
it occupies an exponentially vanishing fraction of the entire Hilbert space. 
This intermediate scaling induces the unique spectral statistics around zero energy, 
as studied in Secs.~\ref{sec:edge} and \ref{sec:gap}.
Notably, a similarly subextensive number of many-body zero modes also appears in the disorder-free case (see Appendix~\ref{appendix:zero} for details).

The zero modes protected by the chiral index must be distinguished from the additional contributions arising from the isolated states when characterizing the spectral statistics. 
Removing the vacuum and fully occupied states leaves all nonzero eigenenergies unchanged, 
although it modifies the dimensions of the remaining zero-mode space. 
Consequently, the original signed indices $\nu_\pm$ need not coincide with the effective indices entering the random-matrix description of the remaining spectrum. 
We clarify this distinction in Sec.~\ref{sec:edge}.

\section{Bulk spectral signatures of quantum chaos}
    \label{sec:bulk}

We now investigate the bulk spectral correlations of many-body levels away from zero energy. 
This analysis provides a test of quantum chaos based on the symmetry classification developed in Sec.~\ref{sec:symmetry}.
To characterize the bulk spectral statistics, 
we order the positive eigenenergies within the fixed fermion-parity subspace as $E_{n-1} < E_n < E_{n+1}$, 
and consider the level-spacing ratio~\cite{Oganesyan-07, Atas-13}
\begin{equation}
    r_n \coloneqq
    \frac{\min \left( E_{n+1}-E_n,E_n-E_{n-1} \right)}
         {\max \left( E_{n+1}-E_n,E_n-E_{n-1} \right)} \in \left[ 0, 1 \right].
\end{equation}
Unlike the level spacing $E_{n+1} - E_n$ itself, 
the ratio $r_n$ is invariant under the common rescaling of the two adjacent spacings. 
When the mean density of states varies smoothly on the scale of these spacings, 
its local contribution cancels in the ratio, 
allowing direct comparison with the random-matrix statistics without spectral unfolding. 

For the Gaussian orthogonal, unitary, and symplectic ensembles, 
the level-spacing-ratio distributions are accurately approximated by the Wigner-like surmise~\cite{Atas-13}
\begin{equation}
    p_r(r) =
    \frac{1}{Z_\beta}
    \frac{(r+r^2)^\beta}
         {(1+r+r^2)^{1+3\beta/2}}
    \quad \left( 0\le r\le1 \right),
    \label{eq:ratio_distribution}
\end{equation}
where $\beta = 1, 2, 4$ is the Dyson index, and the normalization constants $Z_\beta$ are given as
\begin{equation}
    Z_1=\frac{4}{27},
    \quad
    Z_2=\frac{2\pi}{81\sqrt{3}},
    \quad
    Z_4=\frac{2\pi}{729\sqrt{3}}.
\end{equation}
While Eq.~\eqref{eq:ratio_distribution} is derived from the joint distribution of three eigenvalues in the corresponding Gaussian ensemble,
it provides an accurate approximation to the bulk distribution of large random matrices. 
Its small-$r$ behavior, $p_r(r)\propto r^\beta$, reflects the level repulsion. 
By contrast, uncorrelated levels obey the Poisson statistics $p_r(r)=2/(1+r)^2$, which remains nonzero $p_r(0) = 2$ even at $r=0$.

The classification in Sec.~\ref{sec:symmetry} determines the appropriate bulk spectral statistics.
Specifically, classes BDI and CI correspond to $\beta=1$, 
class AIII to $\beta=2$, 
and classes CII and DIII to $\beta=4$. 
Accordingly, Table~\ref{tab:symmetry} gives
\begin{equation}
    \beta=
    \begin{cases}
        1 & \left[ N\equiv0 \left( \text{mod}~4\right) \right],\\
        2 & \left[ N\equiv1,3 \left( \text{mod}~4\right) \right],\\
        4 & \left[ N\equiv2 \left( \text{mod}~4\right) \right].
    \end{cases}
    \label{eq:bulk_Dyson_index}
\end{equation}
For given $N$, the same Dyson index $\beta$ applies to both fermion-parity subspaces $\left( -1 \right)^{\cal F} = \pm 1$,
even when the two subspaces belong to different chiral symmetry classes. 
The bulk statistics thus probe time-reversal symmetry within each subspace, 
although they fail to distinguish all the five classes.

In the numerical analysis, we first remove the zero modes and retain only the positive eigenenergies. 
Since chiral symmetry yields opposite-sign pairs $\left( E, -E \right)$ of eigenenergies, 
the negative-energy spectrum contains no additional information.
For $N\equiv2\pmod4$, we count each Kramers pair only once before computing level spacings; 
otherwise, the exact degeneracy would introduce zero spacings unrelated to correlations between different energy levels. 
We then evaluate the level-spacing ratios in the central $50\%$ of the ordered positive spectrum, 
with the window defined by the number of different levels rather than by a fixed energy interval. 
This procedure excludes both the special region around zero energy and the outer spectral edge.  
For $N=11,12,13,14$, we average the numerical results over $5000$, $2000$, $1000$, and $200$ disorder realizations, respectively.
We analyze the two fermion-parity subspaces separately.

\begin{figure}[t]
\centering
\includegraphics[width=1.0\linewidth]{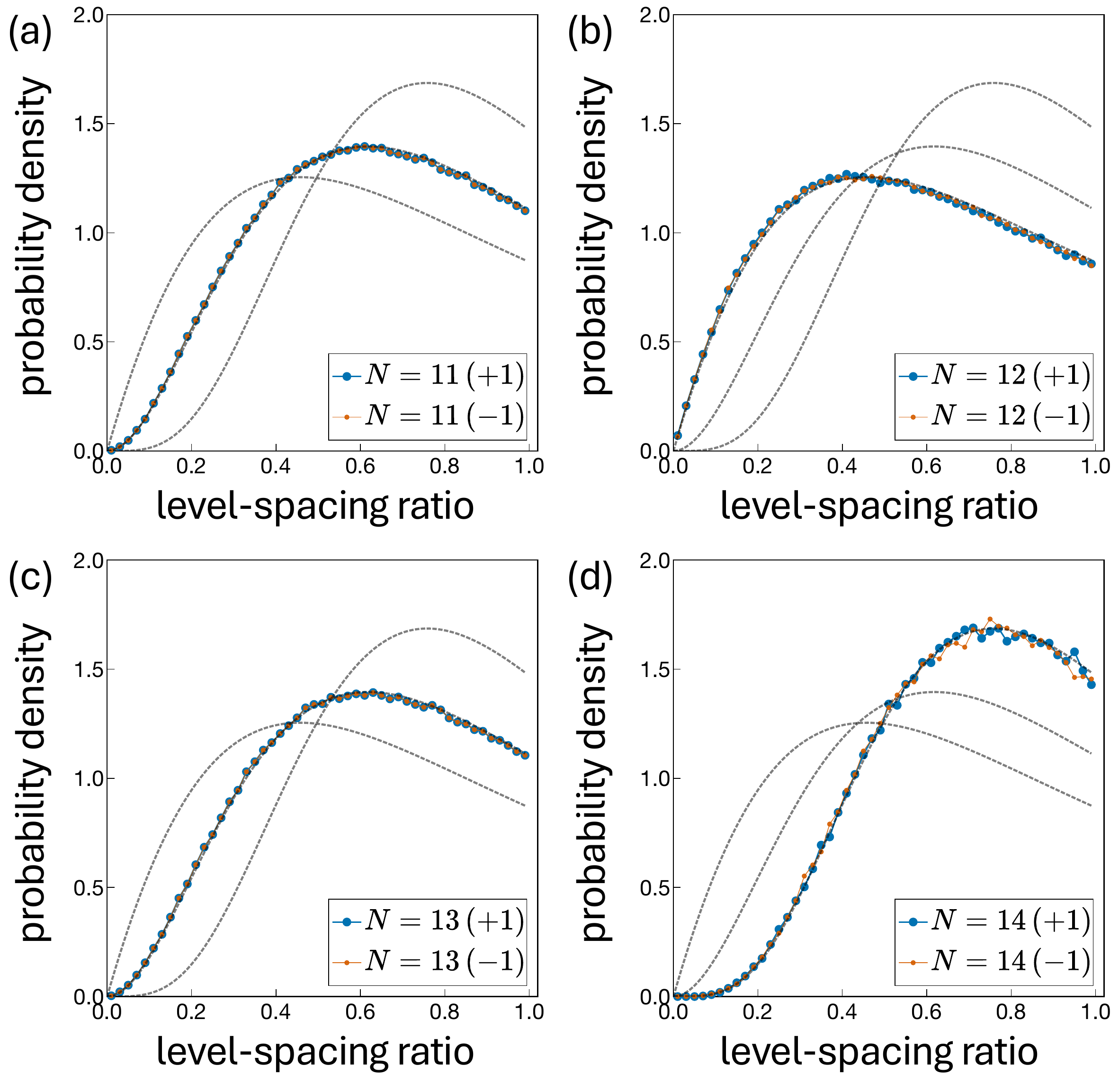} 
\caption{Probability density function of level-spacing ratios for the quantum breakdown model ($J=1$) with (a)~$N=11$, (b)~$N=12$, (c)~$N=13$, and (d)~$N=14$.
The results are averaged over (a)~$5000$, (b)~$2000$, (c)~$1000$, and (d)~$200$ realizations.
All the data are taken from the central 50\% of the positive eigenenergies. 
Blue dots: $\left( -1 \right)^{
\cal F} = +1$.
Orange dots: $\left( -1 \right)^{
\cal F} = -1$.
Black dashed curves: analytical results for small Hermitian random matrices [Eq.~\eqref{eq:ratio_distribution}].}	
    \label{fig:level-spacing-ratio}
\end{figure}

The numerically obtained distributions, shown in Fig.~\ref{fig:level-spacing-ratio}, agree well with the symmetry-dependent random-matrix statistics in Eq.~\eqref{eq:ratio_distribution}. 
For $N=11$ and $13$, both fermion-parity subspaces belong to class AIII, and their distributions follow the $\beta=2$ statistics [see Figs.~\ref{fig:level-spacing-ratio}\,(a) and (c)]. 
The exact coincidence of the distributions for $\left( - 1\right)^{\cal F} = +1$ and $\left( - 1\right)^{\cal F} = -1$ follows from the isospectrality, 
as discussed in Sec.~\ref{sec:symmetry}. 
For $N=12$, 
while the $\left( -1 \right)^{\cal F} = +1$ and $\left( -1 \right)^{\cal F} = -1$ subspaces belong to classes BDI and CI, respectively, 
both exhibit the $\beta=1$ bulk distribution [see Fig.~\ref{fig:level-spacing-ratio}\,(b)]. 
Similarly, for $N=14$, classes DIII and CII yield the same $\beta=4$ bulk statistics [see Fig.~\ref{fig:level-spacing-ratio}\,(d)].
In Table~\ref{tab:data}, we summarize the numerically obtained mean level-spacing ratios $\braket{r}$,
providing a complementary quantitative comparison with the corresponding random-matrix values. 

These results demonstrate quantum chaotic correlations in the spectral bulk, 
despite the presence of exact many-body zero modes. 
The zero-energy degeneracy is thus compatible with the level repulsion in the spectral bulk,
characteristic of quantum chaos. 
At the same time, the agreement between the two fermion-parity distributions for even $N$ shows the limitation of the bulk statistics.
Although they distinguish the three Dyson indices $\beta = 1, 2, 4$, 
they fail to distinguish class BDI from class CI, or class DIII from class CII (see Table~\ref{tab:symmetry}). 
To resolve these additional features encoded by chiral symmetry, 
we turn to the hard-edge spectral statistics in Sec.~\ref{sec:edge}, 
followed by the scaling of the spectral gap and density of states around the zero modes in Sec.~\ref{sec:gap}.

\begin{table*}[t]
	\centering
	\caption{Numerical results for the spectral statistics of the quantum breakdown model.
    For $\braket{r}$, the results are averaged over $5000$ realizations for $N=11$, $2000$ realizations for $N=12$, $1000$ realizations for $N=13$, and $200$ realizations for $N=14$.
    For $\braket{E_{\rm min}^2}/\braket{E_{\rm min}}^2$, the results are averaged over $10000$ realizations for each of $N=11, 12, 13, 14$.
    The uncertainty represents the standard error.
    For comparison, we also present the corresponding results for random matrices that belong to the same symmetry classes as the quantum breakdown model.}
     \begin{tabular}{l|cc} \hline \hline
     ~~Model~~ & ~~$\braket{r}$~~ & ~~$\braket{E_{\rm min}^2}/\braket{E_{\rm min}}^2$~~  \\ \hline
     Quantum breakdown model [$N=11$, $\left( -1 \right)^{\cal F} = \pm 1$] & ~~$0.6000 \pm 0.0003$~~ & ~~$1.00435 \pm 0.00006$~~ \\ 
     Quantum breakdown model [$N=12$, $\left( -1 \right)^{\cal F} = + 1$] & $0.5307 \pm 0.0003$ & $1.00367 \pm 0.00005$ \\ 
     Quantum breakdown model [$N=12$, $\left( -1 \right)^{\cal F} = - 1$] & $0.5311 \pm 0.0003$ & $1.274\pm0.004$ \\ 
     Quantum breakdown model [$N=13$, $\left( -1 \right)^{\cal F} = \pm 1$] & $0.6000 \pm 0.0003$ & $1.00187\pm0.00003$ \\ 
     Quantum breakdown model [$N=14$, $\left( -1 \right)^{\cal F} = + 1$] & $0.6743 \pm 0.0006$ & $1.0831\pm0.0012$\\ 
     Quantum breakdown model [$N=14$, $\left( -1 \right)^{\cal F} = - 1$] & $0.6728 \pm 0.0006$ & $1.000989\pm0.000014$\\ \hline
     Gaussian unitary ensemble (GUE)~\cite{Atas-13} & $0.5996$ & --- \\
     Gaussian orthogonal ensemble (GOE)~\cite{Atas-13} & $0.5307$ & --- \\
     Gaussian symplectic ensemble (GSE)~\cite{Atas-13} & $0.6744$ & --- \\
     Chiral Gaussian unitary ensemble (chGUE) ($\nu = 33$)~\cite{Nishigaki-98, *Damgaard-01} & --- & $1.00414$ \\
     Chiral Gaussian unitary ensemble (chGUE) ($\nu = 65$)~\cite{Nishigaki-98, *Damgaard-01} & --- & $1.00177$ \\
     Chiral Gaussian orthogonal ensemble (chGOE) ($\nu = 66$)~\cite{Nagao-95, Nagao-00} & --- & $1.00350$ \\
     Chiral Gaussian symplectic ensemble (chGSE) ($\nu = 64$)~\cite{Nagao-00, Nishigaki-98, *Damgaard-01} & --- &  $1.000908$ \\
     Class CI~\cite{Nagao-95, Nagao-98} & --- & $1.273$ \\
     Class DIII ($\mathbb{Z}_2$ odd)~\cite{Ivanov-02} & --- & $1.0829$ \\ \hline \hline
    \end{tabular}
	\label{tab:data}
\end{table*}

\section{Hard-edge spectral statistics}
    \label{sec:edge}

\begin{figure}[t]
\centering
\includegraphics[width=1.0\linewidth]{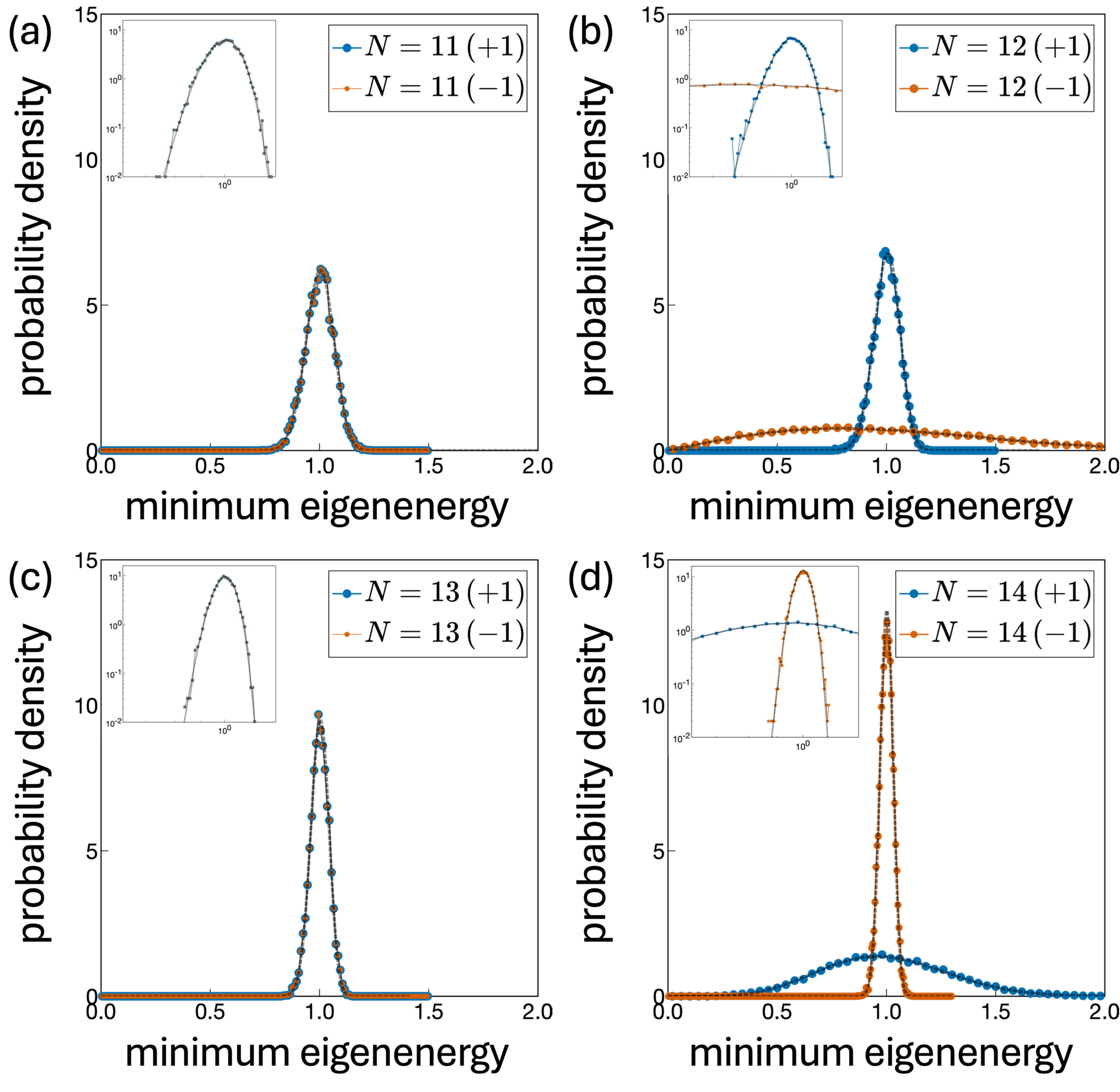} 
\caption{Probability density function of the minimum positive eigenenergy for the quantum breakdown model ($J=1$) with (a)~$N=11$, (b)~$N=12$, (c)~$N=13$, and (d)~$N=14$.
All the results are averaged over $10000$ realizations.
The probability density functions are normalized such that their averages are unity (i.e., $\braket{E_{\rm min}} = 1$).
Blue dots: $\left( -1 \right)^{
\cal F} = +1$.
Orange dots: $\left( -1 \right)^{
\cal F} = -1$.
Black dashed curves: analytical results for the corresponding symmetry classes and chiral indices.
Insets: same distributions on log-log scales.}	
    \label{fig:min-eig}
\end{figure}

While the bulk spectral correlations studied in Sec.~\ref{sec:bulk} distinguish the three Dyson indices,
they do not resolve the full symmetry classification of the quantum breakdown model. 
Around zero energy, chiral symmetry influences the correlations between the nonzero energy levels and the zero-mode subspace. 
We thus investigate the distribution of the minimum positive eigenenergy,
\begin{equation}
E_{\min}\coloneqq\min_{E_n>0}E_n,
    \label{eq:edge_minimum}
\end{equation}
within each fixed subspace of fermion parity. 
These hard-edge statistics provide information beyond that accessible from the  bulk level-spacing ratios,
showing how the same bulk statistics can coexist with the different hard-edge statistics.
In Fig.~\ref{fig:min-eig}, we show the distributions of $E_{\min}/\langle E_{\min}\rangle$, 
obtained from $10000$ disorder realizations for each of $N=11,12,13,14$. 
Each distribution is normalized by its own disorder-averaged minimum eigenenergy $\braket{E_{\rm min}}$, 
so that the shapes of the distributions can be compared independently of the overall energy scale. 
We further compare the dimensionless moment $\langle E_{\min}^{2}\rangle/\langle E_{\min}\rangle^{2}$ in Table~\ref{tab:data}. 
This quantity equals one plus the variance of $E_{\min}/\langle E_{\min}\rangle$ and measures its relative fluctuations.

For the chiral classes (i.e., classes AIII, BDI, and CII), 
the Hamiltonians take the block-off-diagonal form in Eq.~\eqref{eq:zero_chiral_block}, 
with complex, real, and quaternion-real Gaussian off-diagonal blocks $h$, respectively.
The distributions of the smallest positive eigenvalue depend on both the Dyson index $\beta$ and the chiral index $\nu$.
In particular, the probability density exhibits the asymptotic behavior~\cite{Verbaarschot-94, Verbaarschot-00-review}
\begin{equation}
p_{\rm min} \left( E_{\min} \right)\propto E_{\min}^{\alpha}
\quad \left( E_{\min}\to0 \right),
\label{eq:edge_repulsion}
\end{equation}
with
\begin{equation}
\alpha=\beta(\nu+1)-1=
\begin{cases}
\nu & \left( \text{class BDI} \right),\\
2\nu+1 & \left( \text{class AIII} \right),\\
4\nu+3 & \left( \text{class CII} \right).
\end{cases}
    \label{eq:edge_chiral_exponent}
\end{equation}
The exponent $\alpha$ characterizes the strength of level repulsion from the origin. 
On the other hand, class CI is characterized by $\alpha = 1$;
for class DIII, we have $\alpha = 1$ for an even-dimensional antisymmetric off-diagonal block $h$ in Eq.~\eqref{eq:zero_chiral_block} and $\alpha = 5$ for an odd-dimensional block $h$~\cite{AZ-97}.
The hard-edge spectral statistics in Eq.~\eqref{eq:edge_repulsion} have found various applications, for example, in quantum chromodynamics~\cite{Verbaarschot-00-review} and superconductor physics~\cite{AZ-97, Beenakker-review-15}.
While these previous studies focused on the case of $\alpha = \mathcal{O} \left( 1 \right)$,
we here demonstrate that the hard-edge random-matrix statistics remain applicable even when the exponent $\alpha$ becomes subextensive, $\alpha = \mathcal{O}\,( 2^{N/2} )$.

\subsection{Class AIII}

For $N=11$, in the $\left( -1 \right)^{\cal F} = +1$ subspace, the off-diagonal block $h$ in Eq.~\eqref{eq:zero_chiral_block} has dimensions $p = 496$ and $q=528$. 
Removing the isolated vacuum state $\ket{\rm vac}$ reduces the off-diagonal block to dimensions $495\times528$ and gives the effective chiral index
\begin{equation}
    \nu=528-(496-1)=33.
\label{eq:edge_index_N11}
\end{equation}
Accordingly, the spectrum should be compared with the chiral Gaussian unitary ensemble with $\nu=33$~\cite{Verbaarschot-94, Verbaarschot-00-review, Nishigaki-98, *Damgaard-01}, 
rather than $|\nu_+|=32$ or the total zero-mode count $34$. 
Removing the fully occupied state from the $\left( -1 \right)^{\cal F} = -1$ subspace gives the same effective index, 
consistent with the exact isospectrality of the two fermion-parity subspaces.

For $N=13$, in the $\left( -1 \right)^{\cal F} = +1$ subspace, we have $p=2016$ and $q=2080$, 
and the vacuum state $\ket{\rm vac}$ again belongs to the smaller space.
Removing this isolated state reduces the block to dimensions  $2015\times2080$, yielding
\begin{equation}
\nu=2080-(2016-1)=65.
    \label{eq:edge_index_N13}
\end{equation}
The $\left( -1 \right)^{\cal F} = -1$ subspace gives the same result after removing its isolated fully occupied state.
Thus, $N=11$ and $N=13$ realize the same symmetry class but different effective chiral indices.

We show the probability distributions of the minimum positive eigenenergy in Figs.~\ref{fig:min-eig}\,(a) and \ref{fig:min-eig}(c), 
which are well described by the random-matrix statistics of the two chiral Gaussian unitary ensembles with $\nu = 33, 65$. 
Increasing the chiral index from $33$ to $65$ produces a distribution that is more sharply concentrated around its normalized mean. 
Correspondingly, we numerically obtain $\langle E_{\min}^{2}\rangle/\langle E_{\min}\rangle^{2}=1.00435\pm0.00006$ and $1.00187\pm0.00003$, 
in agreement with the reference values $1.00414$ and $1.00177$, respectively (see Table~\ref{tab:data}).  
This distinction illustrates the additional spectral information encoded by the chiral index at the hard edge.

\subsection{Classes BDI and CI}

For $N=12$, the $\left( -1 \right)^{\cal F} = +1$ subspace belongs to class BDI and has $p=992$ and $q=1056$. 
Both $\ket{\rm vac}$ and $\ket{\rm full}$ lie within the smaller space, 
and removing these isolated states reduces the off-diagonal block to a $990\times1056$ real matrix.
The effective chiral index is thus obtained as
\begin{equation}
\nu=1056-(992-2)=66.
    \label{eq:edge_index_N12}
\end{equation}
In Fig.~\ref{fig:min-eig}\,(b),
we show the probability distribution of the minimum positive eigenenergy,
which is narrowly concentrated and agrees well with the random-matrix statistics of the chiral Gaussian orthogonal ensemble with $\nu = 66$~\cite{Nagao-95, Nagao-00}. 
The numerically obtained moment ratio $1.00367\pm0.00005$ is close to the random-matrix value $1.00350$ (see Table~\ref{tab:data}).

By contrast, the $\left( -1 \right)^{\cal F} = -1$ subspace belongs to class CI. 
The corresponding off-diagonal block $h$ is a $1024 \times 1024$ square matrix,
and neither isolated state is present.
Moreover, $h$ is complex symmetric,
giving rise to the hard-edge statistics distinct from those of class BDI,
despite the common bulk Dyson index $\beta = 1$.
For the Gaussian ensemble in class CI, 
the smallest positive eigenvalue conforms to the Rayleigh distribution~\cite{Nagao-95, Nagao-98},
\begin{equation}
p_{\min} \left( E_{\min} \right) =
\frac{\pi E_{\min}}{2\langle E_{\min}\rangle^2}
\exp\left(-\frac{\pi E_{\min}^2}{4\langle E_{\min}\rangle^2}\right),
    \label{eq:edge_CI_distribution}
\end{equation}
which satisfies
\begin{equation}
\frac{\langle E_{\min}^2\rangle}{\langle E_{\min}\rangle^2}
=\frac{4}{\pi}=1.27324\ldots.
    \label{eq:edge_CI_moment}
\end{equation}
The $\left( -1 \right)^{\cal F} = -1$ distribution shown in Fig.~\ref{fig:min-eig}\,(b),
with the numerically obtained moment ratio $1.274\pm0.004$,
agrees well with this broader distribution. 
Thus, although the two fermion-parity subspaces exhibit the same bulk spectral statistics,
their small positive eigenenergies display the distinct relative fluctuations.

\subsection{Classes DIII and CII}

For $N=14$, the $\left( -1 \right)^{\cal F} = +1$ subspace belongs to class DIII, 
and the corresponding off-diagonal block $h$ has dimensions $p=q=4096$. 
After removing the vacuum and fully occupied states, 
the block reduces to a $4095\times4095$ complex antisymmetric matrix. 
In Fig.~\ref{fig:min-eig}\,(d),
we show the probability distribution of the minimum positive eigenenergy,
which agrees well with the random-matrix statistics for class DIII in the odd $\mathbb{Z}_2$ sector~\cite{Ivanov-02}.
The numerically obtained moment ratio $1.0831\pm0.0012$ is also consistent with the random-matrix value $1.0829$ (see Table~\ref{tab:data}).

The $\left( -1 \right)^{\cal F} = -1$ subspace belongs to class CII and contains neither isolated state. 
The corresponding off-diagonal block is a $4032 \times 4160$ matrix, 
yielding the effective chiral index
\begin{equation}
\nu= \frac{1}{2} \left( 4160-4032 \right) =64,
    \label{eq:edge_index_N14}
\end{equation}
where the factor of $1/2$ arises from the Kramers degeneracy.
The probability distribution shown in Fig.~\ref{fig:min-eig}\,(d) is strongly concentrated around its normalized mean,
which contrasts with that for $\left( -1 \right)^{\cal F} = +1$ (i.e., class DIII).
The numerically obtained moment ratio is $1.000989\pm0.000014$, 
close to the random-matrix value $1.000908$ for the chiral Gaussian symplectic ensemble (see Table~\ref{tab:data})~\cite{Nagao-00, Nishigaki-98, *Damgaard-01}. 
This provides another distinction between the two symmetry classes that is invisible in their common $\beta=4$ bulk statistics.

\section{Spectral gap around many-body zero modes}
    \label{sec:gap}

While the hard-edge statistics studied in Sec.~\ref{sec:edge} characterize the relative fluctuations of the first nonzero level, 
their normalization eliminates information about its absolute energy scale. 
We now investigate the separation between the many-body zero modes and the nonzero spectrum,
as well as how the density of states grows beyond this gap region.
In the quantum breakdown model with a  large chiral index, 
we demonstrate that the exponentially large number of zero modes gives rise to the distinct scaling behavior of three characteristic energy scales---spectral gap $\Delta$, mean level spacing $\delta$, and spectral width $W$.

\begin{figure}[t]
\centering
\includegraphics[width=1.0\linewidth]{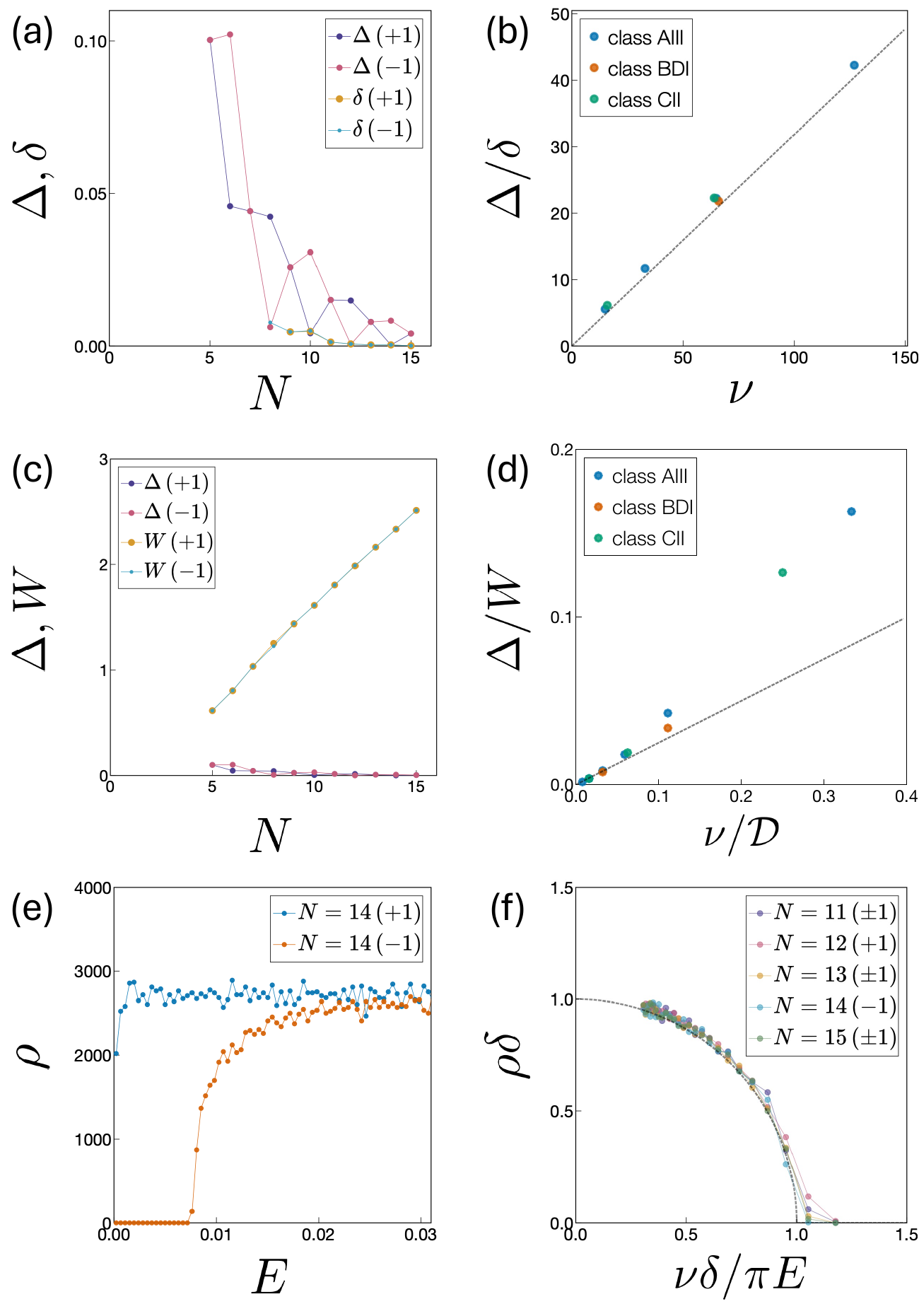} 
\caption{Energy gap scaling of the quantum breakdown model ($J=1$).
The data points are averaged over $1000$ realizations for $N \leq 12$, $500$ realizations for $N=13$, $200$ realizations for $N=14$, and $100$ realizations for $N=15$.
(a)~Energy gap $\Delta$ and mean level spacing $\delta$ as functions of $N$.
(b)~$\Delta/\delta$ as a function of $\nu$.
Black dashed line: $\Delta/\delta = \nu/\pi$ [Eq.~\eqref{eq:gap_inner_scale}].
(c)~$\Delta$ and spectral width $W$ as functions of $N$.
(d)~$\Delta/W$ as a function of $\nu/\mathcal{D}$.
Black dashed line: $\Delta/W = \nu/4\mathcal{D}$ [Eq.~\eqref{eq:gap_inner_edge_width}].
(e)~Density $\rho$ of states as a function of energy $E$ for $N=14$.
(f)~$\rho \delta$ as a function of $\nu\delta/\pi E$.
Black dashed curve: $\rho\delta = \sqrt{1-\left( \nu\delta/\pi E\right)^2}$ [Eq.~\eqref{eq:gap_density_collapse}].}	
    \label{fig:gap}
\end{figure}

\subsection{Spectral scales}

Basic quantities characterizing the spectrum include the gap $\Delta$ and the spectral width $W$,
\begin{equation}
\Delta\coloneqq\langle E_{\min}\rangle,
\quad
W\coloneqq2\langle E_{\max}\rangle,
\label{eq:gap_scales}
\end{equation}
where $E_{\min}$ is the smallest positive eigenenergy defined in Eq.~\eqref{eq:edge_minimum}, 
and $E_{\max} > 0$ is the largest eigenenergy. 
The factor of two in the definition of $W$ follows from chiral symmetry. 
To determine the local mean level spacing $\delta$ independently of $\Delta$, 
we count the different positive levels within a fixed energy window outside the gap region. 
Specifically, using the root-mean-square spectral width
\begin{equation}
w\coloneqq
\sqrt{\frac{\langle\operatorname{Tr}H^2\rangle}{2^{N-1}}}
=
\frac{J\sqrt{(N-1)(N-2)(N-3)}}{4\sqrt{2}\,N}
\label{eq:gap_rms_width}
\end{equation}
to specify this window, we introduce
\begin{equation}
\delta\coloneqq
\frac{0.10w}
{\left\langle\#\{n:0.15w\le E_n<0.25w\}\right\rangle}.
    \label{eq:gap_background_spacing}
\end{equation}

\subsection{Gap scaling}

A simple random-matrix argument explains the relation between the spectral gap $\Delta$ and the chiral index $\nu$.
Let us consider the chiral Gaussian ensemble 
in which the off-diagonal block $h$ has dimensions $p\times q$ [see Eq.~\eqref{eq:zero_chiral_block}], 
with
\begin{equation}
\mathcal{D} = p+q,
\qquad
\nu = p-q.
\end{equation}
As discussed in Sec.~\ref{sec:zero}, the positive eigenenergies are the singular values of $h$.
According to the Marchenko-Pastur law~\cite{Marchenko-Pastur-67}, 
the singular-value spectrum has inner and outer edges proportional to $\sqrt{p}-\sqrt{q}$ and $\sqrt{p}+\sqrt{q}$, respectively, 
where we take $p\ge q$ without loss of generality.
It then follows that $\Delta$ and $W$ satisfy
\begin{equation}
\frac{\Delta}{W}
=
\frac{\sqrt{p}-\sqrt{q}}
{2\left(\sqrt{p}+\sqrt{q}\right)}
\simeq
\frac{\nu}{4\mathcal{D}}
\quad
\left( \nu\ll \mathcal{D} \right).
\label{eq:gap_inner_edge_width}
\end{equation}

Away from both inner and outer spectral edges,
the corresponding Marchenko-Pastur density becomes approximately constant.
Then, the local mean level spacing $\delta$ is given by
\begin{equation}
\frac{1}{\delta}
\simeq
\frac{4\mathcal{D}}{\pi W}.
\label{eq:gap_mean_spacing_rmt}
\end{equation}
Combining Eqs.~\eqref{eq:gap_inner_edge_width} and \eqref{eq:gap_mean_spacing_rmt}, we obtain
\begin{equation}
\Delta
\simeq
\frac{\nu\delta}{\pi}.
\label{eq:gap_inner_scale}
\end{equation}
Thus, a large chiral index $\nu \gg 1$ pushes the nonzero spectrum away from zero energy, on the scale of the local level spacing.

In Fig.~\ref{fig:gap}\,(a), 
we show in the quantum breakdown model that the gap $\Delta$ is substantially larger than the local mean level spacing $\delta$ in the subspaces with large $\nu$.
The contrast between the different fermion-parity subspaces is particularly pronounced for even $N$. 
For $N=12$, $\Delta/\delta$ is approximately $21.8$ in class BDI but only $0.79$ in class CI. 
For $N=14$, it is approximately $22.3$ in class CII but $0.89$ in class DIII. 
Hence, the subspace with the larger gap switches from even parity for $N\equiv0\pmod4$ to odd parity for $N\equiv2\pmod4$, 
following the chiral indices. 
In Fig.~\ref{fig:gap}\,(b),
we also show that $\Delta/\delta$ grows linearly with the chiral index $\nu$ across the chiral symmetry classes, 
confirming the random-matrix relation in Eq.~\eqref{eq:gap_inner_scale}.

We further investigate the scaling behavior of the gap $\Delta$ relative to the full spectral width $W$.
Since the chiral index $\nu$ grows as $2^{N/2}$ whereas the Hilbert-space dimension $\mathcal{D}$ grows as $2^N$, 
Eq.~\eqref{eq:gap_inner_edge_width} shows that $\Delta/W$ decreases exponentially with $N$. 
The numerical results in Figs.~\ref{fig:gap}\,(c) and (d) support this exponential separation of energy scales. 
We also confirm the random-matrix scaling in Eq.~\eqref{eq:gap_inner_edge_width} for $\nu/\mathcal{D} \lesssim 0.1$ across the different chiral classes,
with some deviations appearing for larger $\nu/\mathcal{D}$.

\subsection{Scaling of the density of states}

The redistribution of nonzero energy levels can also be observed directly in the density of states. 
We define the density of states as
\begin{equation}
\rho \left( E \right) \coloneqq
\Braket{
\sum_{n}\delta \left( E-E_n \right)
}.
    \label{eq:gap_density}
\end{equation}
Here, $\rho$ is not the probability density normalized to unit integral but rather the density that counts the number of energy levels per unit energy. 
In Fig.~\ref{fig:gap}\,(e), we compare $\rho$ for the two fermion-parity subspaces at $N=14$.
Despite the similar densities away from the origin, 
the odd-parity spectrum (class CII) exhibits a pronounced gap, 
whereas the even-parity spectrum (class DIII) extends much closer to zero energy.
In addition to the hard-edge spectral statistics in Sec.~\ref{sec:edge},
this contrast demonstrates how the nontrivial chiral index reorganizes the spectrum near zero energy without changing the bulk spectral correlations.

For a large chiral index $\nu \gg 1$, the Marchenko-Pastur density reduces to~\cite{Marchenko-Pastur-67}
\begin{equation}
\rho \left( E \right)\simeq
\begin{cases}
0 & \left( 0 < E <\Delta \right), \\
\displaystyle
\frac{1}{\delta}\sqrt{1-\left(\frac{\Delta}{E}\right)^2}
& \left( \Delta \leq E \ll W/2 \right).
\end{cases}
    \label{eq:gap_density_profile}
\end{equation}
In Fig.~\ref{fig:gap}\,(f), 
we confirm this scaling behavior in the quantum breakdown model.
Specifically, we plot $\rho \delta$ as a function of $\nu \delta/\pi E$ and find that all the numerical results across the  chiral symmetry classes collapse onto the common scaling curve
\begin{equation}
\rho \delta \simeq
\sqrt{1-\left(\frac{\nu\delta}{\pi E}\right)^2}
\quad \left( 0<\frac{\nu\delta}{\pi E}\le1 \right).
    \label{eq:gap_density_collapse}
\end{equation}
Although finite-size deviations remain near the spectral onset $\nu\delta/\pi E \simeq 1$,
the common density profile is clearly observed across different $N$ and symmetry classes.

\section{Discussion}
    \label{sec:conclusion}

We have established the $\mathbb{Z}_4$ symmetry classification of the quantum breakdown model, 
as summarized in Table~\ref{tab:symmetry},
and illustrated how internal symmetry enriches quantum chaos beyond the bulk spectral statistics.
A unique feature of the quantum breakdown model is the emergence of chiral symmetry and a concomitant subextensive number of many-body zero modes.
Although these zero modes occupy a vanishing fraction of the entire Hilbert space, 
the exponentially growing chiral index reorganizes the surrounding nonzero spectrum over energy scales much larger than the local mean level spacing.
We have demonstrated that the exact degeneracy of exponentially many zero modes coexists with random-matrix universality in quantum chaos enriched by chiral symmetry.
While we have focused on a zero-dimensional version of the quantum breakdown model,
the internal symmetry classification developed in this work should also be relevant to higher-dimensional counterparts and related variants,
although additional spatial structure can give rise to further symmetry.

The dynamical consequences of this coexistence remain to be explored.
The spectral form factor and operator growth can reveal how the zero-mode sector and its separation from the nonzero energy levels influence scrambling and response.
Developing a controlled large-$N$ description should also provide a route toward exploring possible connections with gravity.
Another important open question is the possible relation between our symmetry classification and symmetry-protected topological phases.
The symmetry classification of the SYK model can be identified with the anomalous symmetry action at boundaries of $\left( 1+1 \right)$-dimensional topological phases protected by appropriate symmetry~\cite{Fidkowski-Kitaev-10, *Fidkowski-Kitaev-11, Turner-11, You-17}.
It remains unclear whether the $\mathbb{Z}_4$ symmetry classification of the quantum breakdown model developed in this work enables an analogous topological interpretation.

\medskip
\begingroup
\renewcommand{\addcontentsline}[3]{}
\begin{acknowledgments}
K.K. thanks Akira Furusaki and Ken Shiozaki for helpful discussion.
K.K. thanks the Yukawa Institute for Theoretical Physics at Kyoto University,
where this work was partially completed during the workshop ``Localisation 2026" (YITP-W-26-10).
K.K. is supported by JSPS KAKENHI Grants 
No.~JP26H02015, No.~JP26K06970, and No.~JP26K17046, 
and JST FOREST Program Grant No.~JPMJFR256P.
K.G. is supported by Forefront Physics and Mathematics Program to Drive Transformation (FoPM), 
a World-leading Innovative Graduate Study (WINGS) Program,
the University of Tokyo. 
H.K. is supported by JSPS KAKENHI Grants No.~JP23K25783 and No.~JP23K25790.
\end{acknowledgments}
\endgroup

\appendix
    
\section{
Disorder-free quantum breakdown model}
    \label{appendix:zero}


\begin{table*}[t]
	\centering
	\caption{Number of many-body zero modes in the disorder-free quantum breakdown model obtained from the numerical diagonalization.
   The numbers different from the disordered counterpart in Table~\ref{tab:zero-v2} are highlighted in bold [i.e., $N=6, 8, 10, 14, 18$ with $\left( -1 \right)^{\cal F} = +1$].
    For $\left( -1 \right)^{\cal F} = +1$, the total number of zero modes is decomposed into the momentum-resolved counts for $\momentum = 0, 1, \cdots, N-1$.
    For the nonzero entries with $\left( -1 \right)^{\cal F} = -1$, the momentum-resolved zero-mode counts coincide with those for $\left( -1 \right)^{\cal F} = +1$.}
    \label{tab:zero-disorderfree}
    \footnotesize
     \begin{tabular}{c|lc} \hline \hline
     ~~$N$~~ & ~~$\left( -1 \right)^{\cal F} = +1$~~ & $\left( -1 \right)^{\cal F} = -1$  \\ \hline
     $4$ & ~$8 = 2+2+2+2$ & $0$ \\ 
     $5$ & ~$6 = 2+1+1+1+1$ & $6$ \\ 
     $6$ & ~${\bf 8} = 2+1+1+2+1+1$ & $8$ \\ 
     $7$ & ~$8 = 2+1+1+1+1+1+1$ & $8$ \\     
     $8$ & ~${\bf 24} = 4+2+4+2+4+2+4+2$ & $0$ \\ 
     $9$ & ~$16 = 2+2+2+1+2+2+1+2+2$ & $16$ \\ 
     $10$ & ~${\bf 32} = 4+3+3+3+3+4+3+3+3+3$ & $32$ \\ 
     $11$ & ~$34 = 4+3+3+3+3+3+3+3+3+3+3$ & $34$ \\ 
     $12$ & ~$68 = 2+10+0+12+0+10+2+10+0+12+0+10$ & $0$ \\ 
     $13$ & ~$66 = 6+5+5+5+5+5+5+5+5+5+5+5+5$ & $66$ \\ 
     $14$ & ~${\bf 128} = 10+9+9+9+9+9+9+10+9+9+9+9+9+9$~ & $128$ \\  
     $15$ & ~$128=8+9+9+8+9+8+8+9+9+8+8+9+8+9+9$~ & $128$ \\  
     $16$ & ~$256 = 32+0+32+0+32+0+32+0+32+0+32+0+32+0+32+0$~ & $0$ \\ 
     $17$ & ~$256 = 16+15+15+15+15+15+15+15+15+15+15+15+15+15+15+15+15$~ & $256$ \\ 
     $18$ & ~${\bf 512} = 30+28+28+29+28+28+29+28+28+30+28+28+29+28+28+29+28+28$~ & $512$ \\ 
     $19$ & ~$514 = 28+27+27+27+27+27+27+27+27+27+27+27+27+27+27+27+27+27+27$~ & $514$ \\ 
     $20$ & ~$1028 = 2+102+0+102+0+104+0+102+0+102+2+102+0+102+0+104+0+102+0+102$~ & $0$ \\ 
     $21$ & ~$1026 = 50+49+49+48+49+49+48+50+49+48+49+49+48+49+50+48+49+49+48+49+49$~ & $1026$ 
     \\ \hline \hline
    \end{tabular}
\end{table*}

\begin{table}[t]
	\centering
	\caption{Momentum-resolved rank deficiency in the disorder-free quantum breakdown model for $\left( -1 \right)^{\cal F} = +1$.}
     \begin{tabular}{cccc} \hline \hline
     ~~$N$~~ & ~~$\momentum$~~ & ~~$\left( p_\momentum, q_\momentum \right)$~~ & $\operatorname{rank} h_\momentum$  \\ \hline
     $4$ & $0, 2$ & $\left( 1, 1 \right)$ & $0$ \\ 
     $5$ & $0$ & $\left( 2, 2 \right)$ & $1$ \\ 
     $8$ & ~~$1, 3, 5, 7$~~ & $\left( 8, 8 \right)$ & $7$ \\
     $11$ & $0$ & $\left( 46, 48 \right)$ & $45$ \\
     $12$ & $0, 6$ & $\left( 86, 86 \right)$ & $85$ \\
     $13$ & $0$ & $\left( 156, 160 \right)$ & $155$ \\
     $19$ & $0$ & $\left( 6886, 6912 \right)$ & $6885$ \\
     $20$ & $0, 10$ & $\left( 13108, 13108 \right)$ & $13107$ \\
     ~~$21$~~ & $0$ & ~~$\left( 24946, 24994 \right)$~~ & ~~$24945$~~ \\ \hline \hline
    \end{tabular}
	\label{tab:zero-rank}
\end{table}

For comparison, we here consider the disorder-free breakdown model
\begin{equation}
H =\frac{K}{N}\sum_{i=1}^{N}
\sum_{\substack{1\le j<k<l\le N\\j,k,l\ne i}} \left(
c_i^\dagger c_jc_kc_l
+ 
c_l^\dagger c_k^\dagger c_j^\dagger c_i
\right)
    \label{eq:breakdown_model_disorderfree}
\end{equation}
with the uniform real coupling $K \in \mathbb{R}$.
In Table~\ref{tab:zero-disorderfree}, we present the numbers of many-body zero modes in this disorder-free model,
which are consistent with the subextensive scaling in Eq.~\eqref{eq:zero_exponential_scaling}.
Nevertheless, the zero-mode counts for $N=6, 8, 10, 14, 18$ with $\left( - 1 \right)^{\cal F} = +1$ deviate from the disordered counterparts in Table~\ref{tab:zero-v2}.
Notably, the zero-mode counts in Table~\ref{tab:zero-disorderfree} also differ from those in Ref.~\cite{Guan-Katsura-26},
where zero modes occupy a finite fraction of the  Hilbert space [i.e., $\mathcal{D}_{\rm zero} = \mathcal{O}\,( 2^N )$].
This difference originates from the additional constraint $j, k, l \neq i$ imposed in the present Hamiltonian.

A crucial distinction from the disordered case is the emergence of translation symmetry,
\begin{equation}
    THT^{-1} = H, \quad Tc_iT^{-1} = c_{i+1}
\end{equation}
with $c_{N+1} \coloneqq c_1$.
Consequently, the Hamiltonian is block diagonalized according to the eigenvalues $e^{2\pi\ii \momentum/N}$ of $T$ ($\momentum=0, 1, 2, \cdots, N-1$),
and the zero-mode counting in  Eq.~\eqref{eq:zero_rank_count} must be carried out separately within each translation subspace.
Below, we clarify this counting for $N=6, 8, 10, 14, 18$ with $\left( -1 \right)^{\cal F} = +1$.
Similar zero-mode counting was also discussed in Refs.~\cite{Turner-18, Sanada-23}.

\begin{figure}[t]
\centering
\includegraphics[width=1.0\linewidth]{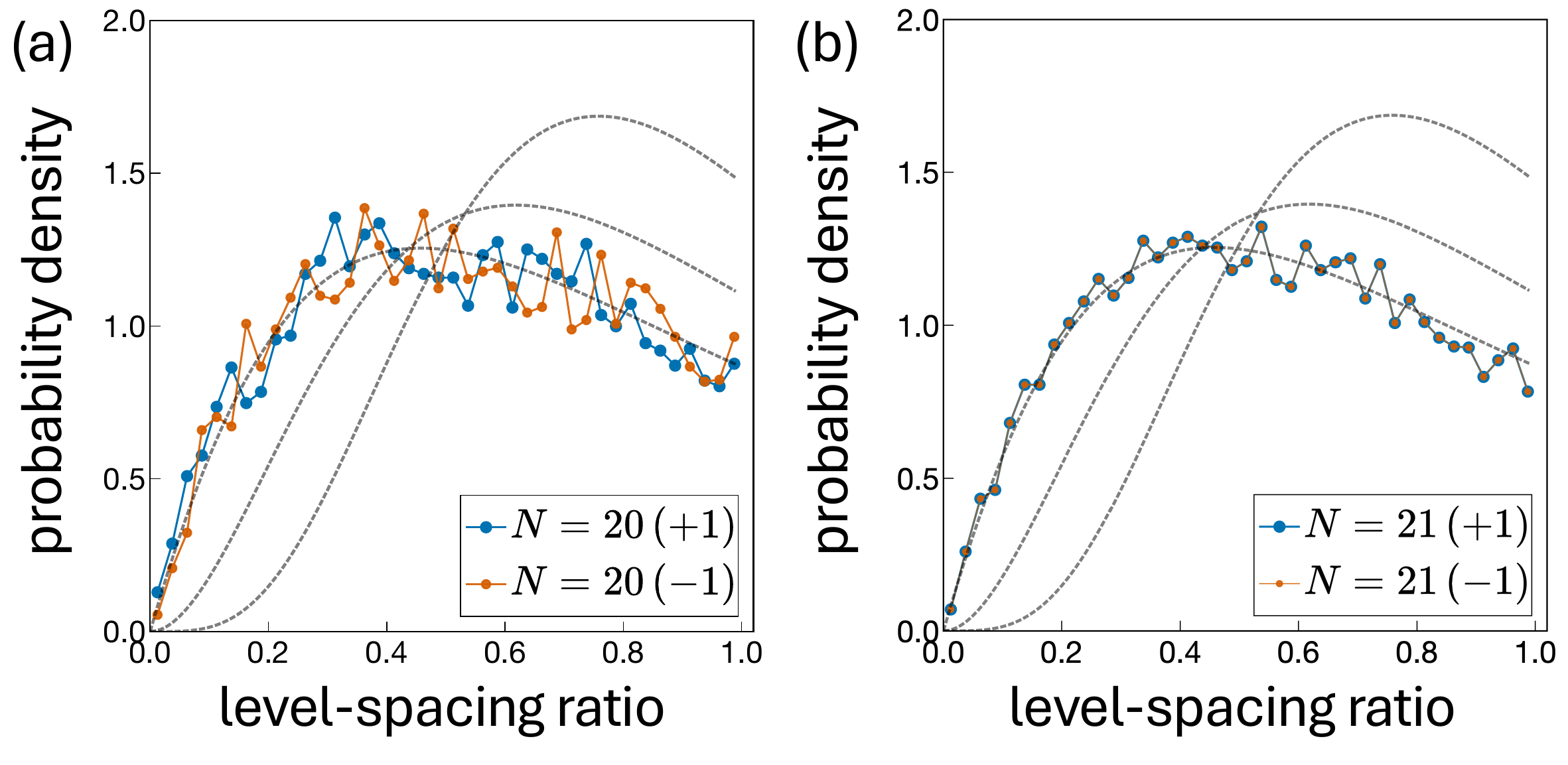}
\caption{Probability density function of level-spacing ratios for the disorder-free quantum breakdown model with (a)~$N=20$ and (b)~$N=21$ ($\momentum=1$).
All the data are taken from the central 50\% of the positive eigenenergies.
(a)~$\braket{r} \simeq 0.5272$ for $\left( -1 \right)^{\cal F} = +1$ and $\braket{r} \simeq 0.5333$ for $\left( -1 \right)^{\cal F} = -1$;
(b)~$\braket{r} \simeq 0.5302$ for $\left( -1 \right)^{\cal F} = \pm 1$.
Blue dots: $\left( -1 \right)^{\cal F} = +1$.
Orange dots: $\left( -1 \right)^{\cal F} = -1$.
Black dashed curves: analytical results for small Hermitian random matrices with $\beta=1,2,4$ [Eq.~\eqref{eq:ratio_distribution}].}
\label{fig:level-spacing-ratio-disorderfree}
\end{figure}

We also investigate the level-spacing-ratio distributions for $N=16, 17, \cdots, 21$ after resolving fermion-parity and translation symmetries.
Even in the absence of disorder in the Hamiltonian,
we find agreement with the Wigner-Dyson random-matrix statistics for $\beta = 1$.
In Fig.~\ref{fig:level-spacing-ratio-disorderfree}, 
we show representative results for $N=20$ and $N=21$ with $k=1$.
It should be noted that the disorder-free model in Eq.~\eqref{eq:breakdown_model_disorderfree} possesses additional symmetry that must be taken into account when analyzing its spectral statistics.
As a representative example,
consider inversion,
\begin{equation}
    I c_i I^{-1} = c_{N+1-i},
\end{equation}
satisfying 
\begin{equation}
    I H I^{-1} = - H.
\end{equation}
While inversion itself anticommutes with the Hamiltonian,
its combination with chiral transformation commutes with the Hamiltonian: 
\begin{equation}
    (\Gamma I)\,H\,(\Gamma I)^{-1} = H.
\end{equation}
Since inversion reverses momentum $k \mapsto -k$,
the combined symmetry $\Gamma I$ is respected within a fixed translation subspace only for $k=0, N/2$.

\subsection{General lower bound for even \texorpdfstring{$N$}{N}}

For arbitrary even $N$, let us resolve the Hilbert space into the eigenspaces of $T^{N/2}$.
Owing to $( T^{N/2} )^2 = 1$,
it decomposes into the two subspaces characterized by $T^{N/2} = +1$ and $T^{N/2} = -1$,
corresponding to even $\momentum$ and odd $\momentum$, respectively.
Within these subspaces, 
the chiral indices $\nu_{+}^{({\rm even/odd})}$ are obtained as
\begin{align}
    \nu_{+}^{({\rm even/odd})} &= \mathrm{Tr}_{\left( -1 \right)^{\cal F} = +1} \left[ \left( \frac{1\pm T^{N/2}}{2} \right) \Gamma \right] \nonumber \\
    &= \frac{1}{2} \left( \nu_{+} \pm 2^{N/2} \right),
\end{align}
where $\nu_{+} = \mathrm{Tr}_{\left( -1 \right)^{\cal F} = +1} \Gamma$ is the total chiral index in Eq.~\eqref{eq:zero_chiral_index_even}.
Additionally, 
for an occupation-number state $\ket{\bm n} = \ket{n_1, \cdots, n_N}$ ($n_{i} = 0, 1$), 
since $\ket{\bm n}$ is an eigenstate of $\Gamma = e^{-\ii \pi \mathcal{F}/2}$ and satisfies $T^{N/2} \ket{\bm n} \propto \ket{n_{N/2+1}, \cdots, n_{N}, n_1, \cdots, n_{N/2}}$,
the matrix element $\braket{{\bm n} | T^{N/2} \Gamma | {\bm n}}$ is nonvanishing only for $n_{i} = n_{i+N/2}$ ($i=1, \cdots, N/2$).
Suppose that the number of nonzero $n_i$'s ($i=1, \cdots, N/2$) is $r$.
Then, we have 
\begin{align}
    &\Gamma \ket{n_1, \cdots, n_{N/2}, n_1, \cdots, n_{N/2}} \nonumber \\
    &\qquad = \left( -1 \right)^{r} \ket{n_1, \cdots, n_{N/2}, n_1, \cdots, n_{N/2}}, \\
    &T^{N/2} \ket{n_1, \cdots, n_{N/2}, n_1, \cdots, n_{N/2}} \nonumber \\
    &\qquad = \left( -1 \right)^{r^2} \ket{n_1, \cdots, n_{N/2}, n_1, \cdots, n_{N/2}},
\end{align}
leading to
\begin{align}
    &T^{N/2}\Gamma \ket{n_1, \cdots, n_{N/2}, n_1, \cdots, n_{N/2}} \nonumber \\
    &\qquad = \left( -1 \right)^{r (r+1)} 
    \ket{n_1, \cdots, n_{N/2}, n_1, \cdots, n_{N/2}} \nonumber \\
    &\qquad = \ket{n_1, \cdots, n_{N/2}, n_1, \cdots, n_{N/2}},
\end{align}
and hence
\begin{align}
    \mathrm{Tr}_{\left( -1 \right)^{\cal F} = +1} T^{N/2} \Gamma
    = \sum_{n_1, \cdots, n_{N/2} = 0, 1} = 2^{N/2}.
\end{align}

Importantly, even if the total chiral index vanishes (i.e., $\nu_{+}^{({\rm even})} + \nu_{+}^{({\rm odd})} = 0$),
as in the case of $N \equiv 2$ (mod $4$),
the chiral indices $\nu_{+}^{({\rm even/odd})}$ within the individual $T^{N/2}$ subspaces can remain nonzero and must be treated separately in the presence of translation symmetry.
Consequently, the number of zero modes satisfies
\begin{equation}
    \mathcal{D}_{\rm zero} \geq | \nu_{+}^{({\rm even})} | + | \nu_{+}^{({\rm odd})} | = 2^{N/2}.
        \label{eq:lowerbound_disorderfree}
\end{equation}
This lower bound is saturated for $N=6, 10, 14, 16, 18$, 
as shown in Table~\ref{tab:zero-disorderfree}.

\subsection{\texorpdfstring{$N=8$}{N=8}}

For $N=8$, the lower bound in Eq.~\eqref{eq:lowerbound_disorderfree} gives $16$,
which is smaller than the actual number of zero modes (i.e., $24$ in Table~\ref{tab:zero-disorderfree}).
We thus need to resolve the Hilbert space into all translation subspaces.
Let $D_{n, \momentum}$ denote the number of $n$-particle configurations with total momentum $2\pi \momentum/N$, 
defined explicitly as
\begin{align}
    &D_{n, \momentum} \coloneqq \# \{ 0 \leq l_1 < \cdots < l_n \leq N-1; \nonumber \\
    &\qquad\qquad\qquad\qquad l_1 + \cdots + l_n \equiv \momentum\,\left( {\rm mod}~N \right) \}.
\end{align}
Then, within an individual translation subspace having $\left( -1 \right)^{\cal F} = +1$,
the dimensions $p_\momentum$ and $q_\momentum$ in Eq.~\eqref{eq:zero_rank_count} are given as
\begin{equation}
    p_\momentum = \sum_{\substack{0\le n\le N\\n\equiv0~\left( \text{mod}\,4 \right)}} D_{n, \momentum}, \quad q_\momentum = \sum_{\substack{0\le n\le N\\n\equiv2~\left( \text{mod}\,4 \right)}} D_{n, \momentum}.
\end{equation}

Specifically, for $N=8$, we find
\begin{itemize}
    \item For $\momentum = 0, 2, 4, 6$, we have $p_\momentum = 10$ and $q_\momentum = 6$, as well as $\operatorname{rank} h_\momentum = 6 = q_\momentum$, leading to $p_\momentum - q_\momentum = 4$ zero modes for each $\momentum$.

    \item For $\momentum = 1, 3, 5, 7$, we have $p_\momentum = q_\momentum = 8$, as well as $\operatorname{rank} h_\momentum = 7 < p_\momentum = q_\momentum$. 
    Accordingly, the rank deficiency gives rise to $2$ zero modes for each $\momentum$.
\end{itemize}
Consequently, the total number of zero modes is given by
\begin{equation}
    \mathcal{D}_{\rm zero} = 4 \times 4 + 2 \times 4 = 24,
\end{equation}
which agrees with the zero-mode count in Table~\ref{tab:zero-disorderfree}.
For $N \leq 21$, we summarize the momentum-resolved rank deficiency in Table~\ref{tab:zero-rank}.

\let\oldaddcontentsline\addcontentsline
\renewcommand{\addcontentsline}[3]{}
\bibliography{ref.bib}

\begin{thebibliography}{63}%
\makeatletter
\providecommand \@ifxundefined [1]{%
 \@ifx{#1\undefined}
}%
\providecommand \@ifnum [1]{%
 \ifnum #1\expandafter \@firstoftwo
 \else \expandafter \@secondoftwo
 \fi
}%
\providecommand \@ifx [1]{%
 \ifx #1\expandafter \@firstoftwo
 \else \expandafter \@secondoftwo
 \fi
}%
\providecommand \natexlab [1]{#1}%
\providecommand \enquote  [1]{``#1''}%
\providecommand \bibnamefont  [1]{#1}%
\providecommand \bibfnamefont [1]{#1}%
\providecommand \citenamefont [1]{#1}%
\providecommand \href@noop [0]{\@secondoftwo}%
\providecommand \href [0]{\begingroup \@sanitize@url \@href}%
\providecommand \@href[1]{\@@startlink{#1}\@@href}%
\providecommand \@@href[1]{\endgroup#1\@@endlink}%
\providecommand \@sanitize@url [0]{\catcode `\\12\catcode `\$12\catcode `\&12\catcode `\#12\catcode `\^12\catcode `\_12\catcode `\%12\relax}%
\providecommand \@@startlink[1]{}%
\providecommand \@@endlink[0]{}%
\providecommand \url  [0]{\begingroup\@sanitize@url \@url }%
\providecommand \@url [1]{\endgroup\@href {#1}{\urlprefix }}%
\providecommand \urlprefix  [0]{URL }%
\providecommand \Eprint [0]{\href }%
\providecommand \doibase [0]{https://doi.org/}%
\providecommand \selectlanguage [0]{\@gobble}%
\providecommand \bibinfo  [0]{\@secondoftwo}%
\providecommand \bibfield  [0]{\@secondoftwo}%
\providecommand \translation [1]{[#1]}%
\providecommand \BibitemOpen [0]{}%
\providecommand \bibitemStop [0]{}%
\providecommand \bibitemNoStop [0]{.\EOS\space}%
\providecommand \EOS [0]{\spacefactor3000\relax}%
\providecommand \BibitemShut  [1]{\csname bibitem#1\endcsname}%
\let\auto@bib@innerbib\@empty
\bibitem [{\citenamefont {Haake}\ \emph {et~al.}(2018)\citenamefont {Haake}, \citenamefont {Gnutzmann},\ and\ \citenamefont {Ku\'{s}}}]{Haake-textbook}%
  \BibitemOpen
  \bibfield  {author} {\bibinfo {author} {\bibfnamefont {F.}~\bibnamefont {Haake}}, \bibinfo {author} {\bibfnamefont {S.}~\bibnamefont {Gnutzmann}},\ and\ \bibinfo {author} {\bibfnamefont {M.}~\bibnamefont {Ku\'{s}}},\ }\href {https://doi.org/10.1007/978-3-319-97580-1} {\emph {\bibinfo {title} {{Quantum Signatures of Chaos}}}}\ (\bibinfo  {publisher} {Springer},\ \bibinfo {address} {Cham, Switzerland},\ \bibinfo {year} {2018})\BibitemShut {NoStop}%
\bibitem [{\citenamefont {Nandkishore}\ and\ \citenamefont {Huse}(2015)}]{Huse-review}%
  \BibitemOpen
  \bibfield  {author} {\bibinfo {author} {\bibfnamefont {R.}~\bibnamefont {Nandkishore}}\ and\ \bibinfo {author} {\bibfnamefont {D.~A.}\ \bibnamefont {Huse}},\ }\bibfield  {title} {\bibinfo {title} {{Many-Body Localization and Thermalization in Quantum Statistical Mechanics}},\ }\href {https://doi.org/10.1146/annurev-conmatphys-031214-014726} {\bibfield  {journal} {\bibinfo  {journal} {Annu. Rev. Condens. Matter Phys.}\ }\textbf {\bibinfo {volume} {6}},\ \bibinfo {pages} {15} (\bibinfo {year} {2015})}\BibitemShut {NoStop}%
\bibitem [{\citenamefont {D'Alessio}\ \emph {et~al.}(2016)\citenamefont {D'Alessio}, \citenamefont {Kafri}, \citenamefont {Polkovnikov},\ and\ \citenamefont {Rigol}}]{Rigol-review}%
  \BibitemOpen
  \bibfield  {author} {\bibinfo {author} {\bibfnamefont {L.}~\bibnamefont {D'Alessio}}, \bibinfo {author} {\bibfnamefont {Y.}~\bibnamefont {Kafri}}, \bibinfo {author} {\bibfnamefont {A.}~\bibnamefont {Polkovnikov}},\ and\ \bibinfo {author} {\bibfnamefont {M.}~\bibnamefont {Rigol}},\ }\bibfield  {title} {\bibinfo {title} {{From quantum chaos and eigenstate thermalization to statistical mechanics and thermodynamics}},\ }\href {https://doi.org/10.1080/00018732.2016.1198134} {\bibfield  {journal} {\bibinfo  {journal} {Adv. Phys.}\ }\textbf {\bibinfo {volume} {65}},\ \bibinfo {pages} {239} (\bibinfo {year} {2016})}\BibitemShut {NoStop}%
\bibitem [{\citenamefont {Berry}\ and\ \citenamefont {Tabor}(1977)}]{Berry-Tabor-77}%
  \BibitemOpen
  \bibfield  {author} {\bibinfo {author} {\bibfnamefont {M.~V.}\ \bibnamefont {Berry}}\ and\ \bibinfo {author} {\bibfnamefont {M.}~\bibnamefont {Tabor}},\ }\bibfield  {title} {\bibinfo {title} {{Level clustering in the regular spectrum}},\ }\href {https://doi.org/10.1098/rspa.1977.0140} {\bibfield  {journal} {\bibinfo  {journal} {Proc. R. Soc. A}\ }\textbf {\bibinfo {volume} {356}},\ \bibinfo {pages} {375} (\bibinfo {year} {1977})}\BibitemShut {NoStop}%
\bibitem [{\citenamefont {Bohigas}\ \emph {et~al.}(1984)\citenamefont {Bohigas}, \citenamefont {Giannoni},\ and\ \citenamefont {Schmit}}]{BGS-84}%
  \BibitemOpen
  \bibfield  {author} {\bibinfo {author} {\bibfnamefont {O.}~\bibnamefont {Bohigas}}, \bibinfo {author} {\bibfnamefont {M.~J.}\ \bibnamefont {Giannoni}},\ and\ \bibinfo {author} {\bibfnamefont {C.}~\bibnamefont {Schmit}},\ }\bibfield  {title} {\bibinfo {title} {{Characterization of Chaotic Quantum Spectra and Universality of Level Fluctuation Laws}},\ }\href {https://doi.org/10.1103/PhysRevLett.52.1} {\bibfield  {journal} {\bibinfo  {journal} {Phys. Rev. Lett.}\ }\textbf {\bibinfo {volume} {52}},\ \bibinfo {pages} {1} (\bibinfo {year} {1984})}\BibitemShut {NoStop}%
\bibitem [{\citenamefont {Wigner}(1951)}]{Wigner-51}%
  \BibitemOpen
  \bibfield  {author} {\bibinfo {author} {\bibfnamefont {E.~P.}\ \bibnamefont {Wigner}},\ }\bibfield  {title} {\bibinfo {title} {{On the statistical distribution of the widths and spacings of nuclear resonance levels}},\ }\href {https://doi.org/10.1017/S0305004100027237} {\bibfield  {journal} {\bibinfo  {journal} {Math. Proc. Cambridge Philos. Soc.}\ }\textbf {\bibinfo {volume} {47}},\ \bibinfo {pages} {790} (\bibinfo {year} {1951})}\BibitemShut {NoStop}%
\bibitem [{\citenamefont {Wigner}(1958)}]{Wigner-58}%
  \BibitemOpen
  \bibfield  {author} {\bibinfo {author} {\bibfnamefont {E.~P.}\ \bibnamefont {Wigner}},\ }\bibfield  {title} {\bibinfo {title} {{On the Distribution of the Roots of Certain Symmetric Matrices}},\ }\href {https://doi.org/10.2307/1970008} {\bibfield  {journal} {\bibinfo  {journal} {Ann. Math.}\ }\textbf {\bibinfo {volume} {67}},\ \bibinfo {pages} {325} (\bibinfo {year} {1958})}\BibitemShut {NoStop}%
\bibitem [{\citenamefont {Dyson}(1962)}]{Dyson-62}%
  \BibitemOpen
  \bibfield  {author} {\bibinfo {author} {\bibfnamefont {F.~J.}\ \bibnamefont {Dyson}},\ }\bibfield  {title} {\bibinfo {title} {{The Threefold Way. Algebraic Structure of Symmetry Groups and Ensembles in Quantum Mechanics}},\ }\href {https://doi.org/10.1063/1.1703863} {\bibfield  {journal} {\bibinfo  {journal} {J. Math. Phys.}\ }\textbf {\bibinfo {volume} {3}},\ \bibinfo {pages} {1199} (\bibinfo {year} {1962})}\BibitemShut {NoStop}%
\bibitem [{\citenamefont {Mehta}(2004)}]{Mehta-textbook}%
  \BibitemOpen
  \bibfield  {author} {\bibinfo {author} {\bibfnamefont {M.~L.}\ \bibnamefont {Mehta}},\ }\href@noop {} {\emph {\bibinfo {title} {{Random Matrices}}}}\ (\bibinfo  {publisher} {Elsevier},\ \bibinfo {address} {Amsterdam},\ \bibinfo {year} {2004})\BibitemShut {NoStop}%
\bibitem [{\citenamefont {Gade}\ and\ \citenamefont {Wegner}(1991)}]{Gade-91}%
  \BibitemOpen
  \bibfield  {author} {\bibinfo {author} {\bibfnamefont {R.}~\bibnamefont {Gade}}\ and\ \bibinfo {author} {\bibfnamefont {F.}~\bibnamefont {Wegner}},\ }\bibfield  {title} {\bibinfo {title} {{The $n = 0$ replica limit of $\mathrm{U} \left( n \right)$ and $\mathrm{U} \left( n \right)/\mathrm{SO} \left( n \right)$ models}},\ }\href {https://doi.org/10.1016/0550-3213(91)90401-I} {\bibfield  {journal} {\bibinfo  {journal} {Nucl. Phys. B}\ }\textbf {\bibinfo {volume} {360}},\ \bibinfo {pages} {213} (\bibinfo {year} {1991})}\BibitemShut {NoStop}%
\bibitem [{\citenamefont {Gade}(1993)}]{Gade-93}%
  \BibitemOpen
  \bibfield  {author} {\bibinfo {author} {\bibfnamefont {R.}~\bibnamefont {Gade}},\ }\bibfield  {title} {\bibinfo {title} {{Anderson localization for sublattice models}},\ }\href {https://doi.org/10.1016/0550-3213(93)90601-K} {\bibfield  {journal} {\bibinfo  {journal} {Nucl. Phys. B}\ }\textbf {\bibinfo {volume} {398}},\ \bibinfo {pages} {499} (\bibinfo {year} {1993})}\BibitemShut {NoStop}%
\bibitem [{\citenamefont {Verbaarschot}(1994)}]{Verbaarschot-94}%
  \BibitemOpen
  \bibfield  {author} {\bibinfo {author} {\bibfnamefont {J.}~\bibnamefont {Verbaarschot}},\ }\bibfield  {title} {\bibinfo {title} {{Spectrum of the QCD Dirac operator and chiral random matrix theory}},\ }\href {https://doi.org/10.1103/PhysRevLett.72.2531} {\bibfield  {journal} {\bibinfo  {journal} {Phys. Rev. Lett.}\ }\textbf {\bibinfo {volume} {72}},\ \bibinfo {pages} {2531} (\bibinfo {year} {1994})}\BibitemShut {NoStop}%
\bibitem [{\citenamefont {Verbaarschot}\ and\ \citenamefont {Wettig}(2000)}]{Verbaarschot-00-review}%
  \BibitemOpen
  \bibfield  {author} {\bibinfo {author} {\bibfnamefont {J.~J.~M.}\ \bibnamefont {Verbaarschot}}\ and\ \bibinfo {author} {\bibfnamefont {T.}~\bibnamefont {Wettig}},\ }\bibfield  {title} {\bibinfo {title} {{Random Matrix Theory and Chiral Symmetry in QCD}},\ }\href {https://doi.org/10.1146/annurev.nucl.50.1.343} {\bibfield  {journal} {\bibinfo  {journal} {Annu. Rev. Nucl. Part. Sci.}\ }\textbf {\bibinfo {volume} {50}},\ \bibinfo {pages} {343} (\bibinfo {year} {2000})}\BibitemShut {NoStop}%
\bibitem [{\citenamefont {Altland}\ and\ \citenamefont {Zirnbauer}(1997)}]{AZ-97}%
  \BibitemOpen
  \bibfield  {author} {\bibinfo {author} {\bibfnamefont {A.}~\bibnamefont {Altland}}\ and\ \bibinfo {author} {\bibfnamefont {M.~R.}\ \bibnamefont {Zirnbauer}},\ }\bibfield  {title} {\bibinfo {title} {{Nonstandard symmetry classes in mesoscopic normal-superconducting hybrid structures}},\ }\href {https://doi.org/10.1103/PhysRevB.55.1142} {\bibfield  {journal} {\bibinfo  {journal} {Phys. Rev. B}\ }\textbf {\bibinfo {volume} {55}},\ \bibinfo {pages} {1142} (\bibinfo {year} {1997})}\BibitemShut {NoStop}%
\bibitem [{\citenamefont {Beenakker}(1997)}]{Beenakker-review-97}%
  \BibitemOpen
  \bibfield  {author} {\bibinfo {author} {\bibfnamefont {C.~W.~J.}\ \bibnamefont {Beenakker}},\ }\bibfield  {title} {\bibinfo {title} {{Random-matrix theory of quantum transport}},\ }\href {https://doi.org/10.1103/RevModPhys.69.731} {\bibfield  {journal} {\bibinfo  {journal} {Rev. Mod. Phys.}\ }\textbf {\bibinfo {volume} {69}},\ \bibinfo {pages} {731} (\bibinfo {year} {1997})}\BibitemShut {NoStop}%
\bibitem [{\citenamefont {Beenakker}(2015)}]{Beenakker-review-15}%
  \BibitemOpen
  \bibfield  {author} {\bibinfo {author} {\bibfnamefont {C.~W.~J.}\ \bibnamefont {Beenakker}},\ }\bibfield  {title} {\bibinfo {title} {{Random-matrix theory of Majorana fermions and topological superconductors}},\ }\href {https://doi.org/10.1103/RevModPhys.87.1037} {\bibfield  {journal} {\bibinfo  {journal} {Rev. Mod. Phys.}\ }\textbf {\bibinfo {volume} {87}},\ \bibinfo {pages} {1037} (\bibinfo {year} {2015})}\BibitemShut {NoStop}%
\bibitem [{\citenamefont {Evers}\ and\ \citenamefont {Mirlin}(2008)}]{Evers-review}%
  \BibitemOpen
  \bibfield  {author} {\bibinfo {author} {\bibfnamefont {F.}~\bibnamefont {Evers}}\ and\ \bibinfo {author} {\bibfnamefont {A.~D.}\ \bibnamefont {Mirlin}},\ }\bibfield  {title} {\bibinfo {title} {{Anderson transitions}},\ }\href {https://doi.org/10.1103/RevModPhys.80.1355} {\bibfield  {journal} {\bibinfo  {journal} {Rev. Mod. Phys.}\ }\textbf {\bibinfo {volume} {80}},\ \bibinfo {pages} {1355} (\bibinfo {year} {2008})}\BibitemShut {NoStop}%
\bibitem [{\citenamefont {Schnyder}\ \emph {et~al.}(2008)\citenamefont {Schnyder}, \citenamefont {Ryu}, \citenamefont {Furusaki},\ and\ \citenamefont {Ludwig}}]{Schnyder-08}%
  \BibitemOpen
  \bibfield  {author} {\bibinfo {author} {\bibfnamefont {A.~P.}\ \bibnamefont {Schnyder}}, \bibinfo {author} {\bibfnamefont {S.}~\bibnamefont {Ryu}}, \bibinfo {author} {\bibfnamefont {A.}~\bibnamefont {Furusaki}},\ and\ \bibinfo {author} {\bibfnamefont {A.~W.~W.}\ \bibnamefont {Ludwig}},\ }\bibfield  {title} {\bibinfo {title} {{Classification of topological insulators and superconductors in three spatial dimensions}},\ }\href {https://doi.org/10.1103/PhysRevB.78.195125} {\bibfield  {journal} {\bibinfo  {journal} {Phys. Rev. B}\ }\textbf {\bibinfo {volume} {78}},\ \bibinfo {pages} {195125} (\bibinfo {year} {2008})}\BibitemShut {NoStop}%
\bibitem [{\citenamefont {Ryu}\ \emph {et~al.}(2010)\citenamefont {Ryu}, \citenamefont {Schnyder}, \citenamefont {Furusaki},\ and\ \citenamefont {Ludwig}}]{Ryu-10}%
  \BibitemOpen
  \bibfield  {author} {\bibinfo {author} {\bibfnamefont {S.}~\bibnamefont {Ryu}}, \bibinfo {author} {\bibfnamefont {A.~P.}\ \bibnamefont {Schnyder}}, \bibinfo {author} {\bibfnamefont {A.}~\bibnamefont {Furusaki}},\ and\ \bibinfo {author} {\bibfnamefont {A.~W.~W.}\ \bibnamefont {Ludwig}},\ }\bibfield  {title} {\bibinfo {title} {{Topological insulators and superconductors: tenfold way and dimensional hierarchy}},\ }\href {https://doi.org/10.1088/1367-2630/12/6/065010} {\bibfield  {journal} {\bibinfo  {journal} {New J. Phys.}\ }\textbf {\bibinfo {volume} {12}},\ \bibinfo {pages} {065010} (\bibinfo {year} {2010})}\BibitemShut {NoStop}%
\bibitem [{\citenamefont {Kitaev}(2009)}]{Kitaev-09}%
  \BibitemOpen
  \bibfield  {author} {\bibinfo {author} {\bibfnamefont {A.}~\bibnamefont {Kitaev}},\ }\bibfield  {title} {\bibinfo {title} {{Periodic table for topological insulators and superconductors}},\ }\href {https://doi.org/10.1063/1.3149495} {\bibfield  {journal} {\bibinfo  {journal} {AIP Conf. Proc.}\ }\textbf {\bibinfo {volume} {1134}},\ \bibinfo {pages} {22} (\bibinfo {year} {2009})}\BibitemShut {NoStop}%
\bibitem [{\citenamefont {Hasan}\ and\ \citenamefont {Kane}(2010)}]{HK-review}%
  \BibitemOpen
  \bibfield  {author} {\bibinfo {author} {\bibfnamefont {M.~Z.}\ \bibnamefont {Hasan}}\ and\ \bibinfo {author} {\bibfnamefont {C.~L.}\ \bibnamefont {Kane}},\ }\bibfield  {title} {\bibinfo {title} {{Colloquium: Topological insulators}},\ }\href {https://doi.org/10.1103/RevModPhys.82.3045} {\bibfield  {journal} {\bibinfo  {journal} {Rev. Mod. Phys.}\ }\textbf {\bibinfo {volume} {82}},\ \bibinfo {pages} {3045} (\bibinfo {year} {2010})}\BibitemShut {NoStop}%
\bibitem [{\citenamefont {Qi}\ and\ \citenamefont {Zhang}(2011)}]{QZ-review}%
  \BibitemOpen
  \bibfield  {author} {\bibinfo {author} {\bibfnamefont {X.-L.}\ \bibnamefont {Qi}}\ and\ \bibinfo {author} {\bibfnamefont {S.-C.}\ \bibnamefont {Zhang}},\ }\bibfield  {title} {\bibinfo {title} {{Topological insulators and superconductors}},\ }\href {https://doi.org/10.1103/RevModPhys.83.1057} {\bibfield  {journal} {\bibinfo  {journal} {Rev. Mod. Phys.}\ }\textbf {\bibinfo {volume} {83}},\ \bibinfo {pages} {1057} (\bibinfo {year} {2011})}\BibitemShut {NoStop}%
\bibitem [{\citenamefont {Chiu}\ \emph {et~al.}(2016)\citenamefont {Chiu}, \citenamefont {Teo}, \citenamefont {Schnyder},\ and\ \citenamefont {Ryu}}]{CTSR-review}%
  \BibitemOpen
  \bibfield  {author} {\bibinfo {author} {\bibfnamefont {C.-K.}\ \bibnamefont {Chiu}}, \bibinfo {author} {\bibfnamefont {J.~C.~Y.}\ \bibnamefont {Teo}}, \bibinfo {author} {\bibfnamefont {A.~P.}\ \bibnamefont {Schnyder}},\ and\ \bibinfo {author} {\bibfnamefont {S.}~\bibnamefont {Ryu}},\ }\bibfield  {title} {\bibinfo {title} {{Classification of topological quantum matter with symmetries}},\ }\href {https://doi.org/10.1103/RevModPhys.88.035005} {\bibfield  {journal} {\bibinfo  {journal} {Rev. Mod. Phys.}\ }\textbf {\bibinfo {volume} {88}},\ \bibinfo {pages} {035005} (\bibinfo {year} {2016})}\BibitemShut {NoStop}%
\bibitem [{\citenamefont {Sachdev}\ and\ \citenamefont {Ye}(1993)}]{Sachdev-Ye-93}%
  \BibitemOpen
  \bibfield  {author} {\bibinfo {author} {\bibfnamefont {S.}~\bibnamefont {Sachdev}}\ and\ \bibinfo {author} {\bibfnamefont {J.}~\bibnamefont {Ye}},\ }\bibfield  {title} {\bibinfo {title} {{Gapless spin-fluid ground state in a random quantum Heisenberg magnet}},\ }\href {https://doi.org/10.1103/PhysRevLett.70.3339} {\bibfield  {journal} {\bibinfo  {journal} {Phys. Rev. Lett.}\ }\textbf {\bibinfo {volume} {70}},\ \bibinfo {pages} {3339} (\bibinfo {year} {1993})}\BibitemShut {NoStop}%
\bibitem [{\citenamefont {Kitaev}(2015)}]{Kitaev-KITP15}%
  \BibitemOpen
  \bibfield  {author} {\bibinfo {author} {\bibfnamefont {A.}~\bibnamefont {Kitaev}},\ }\bibfield  {title} {\bibinfo {title} {{A simple model of quantum holography}}} (\bibinfo {year} {2015}),\ \bibinfo {note} {{KITP Program: Entanglement in Strongly-Correlated Quantum Matter}}\BibitemShut {NoStop}%
\bibitem [{\citenamefont {Sachdev}(2015)}]{Sachdev-15}%
  \BibitemOpen
  \bibfield  {author} {\bibinfo {author} {\bibfnamefont {S.}~\bibnamefont {Sachdev}},\ }\bibfield  {title} {\bibinfo {title} {{Bekenstein-Hawking Entropy and Strange Metals}},\ }\href {https://doi.org/10.1103/PhysRevX.5.041025} {\bibfield  {journal} {\bibinfo  {journal} {Phys. Rev. X}\ }\textbf {\bibinfo {volume} {5}},\ \bibinfo {pages} {041025} (\bibinfo {year} {2015})}\BibitemShut {NoStop}%
\bibitem [{\citenamefont {Polchinski}\ and\ \citenamefont {Rosenhaus}(2016)}]{Polchinski-Rosenhaus-16}%
  \BibitemOpen
  \bibfield  {author} {\bibinfo {author} {\bibfnamefont {J.}~\bibnamefont {Polchinski}}\ and\ \bibinfo {author} {\bibfnamefont {V.}~\bibnamefont {Rosenhaus}},\ }\bibfield  {title} {\bibinfo {title} {{The spectrum in the Sachdev-Ye-Kitaev model}},\ }\href {https://doi.org/10.1007/JHEP04(2016)001} {\bibfield  {journal} {\bibinfo  {journal} {J. High Energ. Phys.}\ }\textbf {\bibinfo {volume} {2016}}\bibinfo  {number} { (4)},\ \bibinfo {pages} {1}}\BibitemShut {NoStop}%
\bibitem [{\citenamefont {Maldacena}\ and\ \citenamefont {Stanford}(2016)}]{Maldacena-Stanford-16}%
  \BibitemOpen
\bibfield  {number} {  }\bibfield  {author} {\bibinfo {author} {\bibfnamefont {J.}~\bibnamefont {Maldacena}}\ and\ \bibinfo {author} {\bibfnamefont {D.}~\bibnamefont {Stanford}},\ }\bibfield  {title} {\bibinfo {title} {{Remarks on the Sachdev-Ye-Kitaev model}},\ }\href {https://doi.org/10.1103/PhysRevD.94.106002} {\bibfield  {journal} {\bibinfo  {journal} {Phys. Rev. D}\ }\textbf {\bibinfo {volume} {94}},\ \bibinfo {pages} {106002} (\bibinfo {year} {2016})}\BibitemShut {NoStop}%
\bibitem [{\citenamefont {Gu}\ \emph {et~al.}(2017)\citenamefont {Gu}, \citenamefont {Qi},\ and\ \citenamefont {Stanford}}]{Gu-17}%
  \BibitemOpen
  \bibfield  {author} {\bibinfo {author} {\bibfnamefont {Y.}~\bibnamefont {Gu}}, \bibinfo {author} {\bibfnamefont {X.-L.}\ \bibnamefont {Qi}},\ and\ \bibinfo {author} {\bibfnamefont {D.}~\bibnamefont {Stanford}},\ }\bibfield  {title} {\bibinfo {title} {{Local criticality, diffusion and chaos in generalized Sachdev-Ye-Kitaev models}},\ }\href {https://doi.org/10.1007/JHEP05(2017)125} {\bibfield  {journal} {\bibinfo  {journal} {J. High Energ. Phys.}\ }\textbf {\bibinfo {volume} {2017}}\bibinfo  {number} { (5)},\ \bibinfo {pages} {125}}\BibitemShut {NoStop}%
\bibitem [{\citenamefont {Fu}\ \emph {et~al.}(2017)\citenamefont {Fu}, \citenamefont {Gaiotto}, \citenamefont {Maldacena},\ and\ \citenamefont {Sachdev}}]{Fu-17}%
  \BibitemOpen
\bibfield  {number} {  }\bibfield  {author} {\bibinfo {author} {\bibfnamefont {W.}~\bibnamefont {Fu}}, \bibinfo {author} {\bibfnamefont {D.}~\bibnamefont {Gaiotto}}, \bibinfo {author} {\bibfnamefont {J.}~\bibnamefont {Maldacena}},\ and\ \bibinfo {author} {\bibfnamefont {S.}~\bibnamefont {Sachdev}},\ }\bibfield  {title} {\bibinfo {title} {{Supersymmetric Sachdev-Ye-Kitaev models}},\ }\href {https://doi.org/10.1103/PhysRevD.95.026009} {\bibfield  {journal} {\bibinfo  {journal} {Phys. Rev. D}\ }\textbf {\bibinfo {volume} {95}},\ \bibinfo {pages} {026009} (\bibinfo {year} {2017})}\BibitemShut {NoStop}%
\bibitem [{\citenamefont {Song}\ \emph {et~al.}(2017)\citenamefont {Song}, \citenamefont {Jian},\ and\ \citenamefont {Balents}}]{Song-17}%
  \BibitemOpen
  \bibfield  {author} {\bibinfo {author} {\bibfnamefont {X.-Y.}\ \bibnamefont {Song}}, \bibinfo {author} {\bibfnamefont {C.-M.}\ \bibnamefont {Jian}},\ and\ \bibinfo {author} {\bibfnamefont {L.}~\bibnamefont {Balents}},\ }\bibfield  {title} {\bibinfo {title} {{Strongly Correlated Metal Built from Sachdev-Ye-Kitaev Models}},\ }\href {https://doi.org/10.1103/PhysRevLett.119.216601} {\bibfield  {journal} {\bibinfo  {journal} {Phys. Rev. Lett.}\ }\textbf {\bibinfo {volume} {119}},\ \bibinfo {pages} {216601} (\bibinfo {year} {2017})}\BibitemShut {NoStop}%
\bibitem [{\citenamefont {Rosenhaus}(2019)}]{Rosenhaus-review}%
  \BibitemOpen
  \bibfield  {author} {\bibinfo {author} {\bibfnamefont {V.}~\bibnamefont {Rosenhaus}},\ }\bibfield  {title} {\bibinfo {title} {{An introduction to the SYK model}},\ }\href {https://doi.org/10.1088/1751-8121/ab2ce1} {\bibfield  {journal} {\bibinfo  {journal} {J. Phys. A}\ }\textbf {\bibinfo {volume} {52}},\ \bibinfo {pages} {323001} (\bibinfo {year} {2019})}\BibitemShut {NoStop}%
\bibitem [{\citenamefont {Chowdhury}\ \emph {et~al.}(2022)\citenamefont {Chowdhury}, \citenamefont {Georges}, \citenamefont {Parcollet},\ and\ \citenamefont {Sachdev}}]{Sachdev-review}%
  \BibitemOpen
  \bibfield  {author} {\bibinfo {author} {\bibfnamefont {D.}~\bibnamefont {Chowdhury}}, \bibinfo {author} {\bibfnamefont {A.}~\bibnamefont {Georges}}, \bibinfo {author} {\bibfnamefont {O.}~\bibnamefont {Parcollet}},\ and\ \bibinfo {author} {\bibfnamefont {S.}~\bibnamefont {Sachdev}},\ }\bibfield  {title} {\bibinfo {title} {{Sachdev-Ye-Kitaev models and beyond: Window into non-Fermi liquids}},\ }\href {https://doi.org/10.1103/RevModPhys.94.035004} {\bibfield  {journal} {\bibinfo  {journal} {Rev. Mod. Phys.}\ }\textbf {\bibinfo {volume} {94}},\ \bibinfo {pages} {035004} (\bibinfo {year} {2022})}\BibitemShut {NoStop}%
\bibitem [{\citenamefont {You}\ \emph {et~al.}(2017)\citenamefont {You}, \citenamefont {Ludwig},\ and\ \citenamefont {Xu}}]{You-17}%
  \BibitemOpen
  \bibfield  {author} {\bibinfo {author} {\bibfnamefont {Y.-Z.}\ \bibnamefont {You}}, \bibinfo {author} {\bibfnamefont {A.~W.~W.}\ \bibnamefont {Ludwig}},\ and\ \bibinfo {author} {\bibfnamefont {C.}~\bibnamefont {Xu}},\ }\bibfield  {title} {\bibinfo {title} {{Sachdev-Ye-Kitaev model and thermalization on the boundary of many-body localized fermionic symmetry-protected topological states}},\ }\href {https://doi.org/10.1103/PhysRevB.95.115150} {\bibfield  {journal} {\bibinfo  {journal} {Phys. Rev. B}\ }\textbf {\bibinfo {volume} {95}},\ \bibinfo {pages} {115150} (\bibinfo {year} {2017})}\BibitemShut {NoStop}%
\bibitem [{\citenamefont {Fu}\ and\ \citenamefont {Sachdev}(2016)}]{Fu-16}%
  \BibitemOpen
  \bibfield  {author} {\bibinfo {author} {\bibfnamefont {W.}~\bibnamefont {Fu}}\ and\ \bibinfo {author} {\bibfnamefont {S.}~\bibnamefont {Sachdev}},\ }\bibfield  {title} {\bibinfo {title} {{Numerical study of fermion and boson models with infinite-range random interactions}},\ }\href {https://doi.org/10.1103/PhysRevB.94.035135} {\bibfield  {journal} {\bibinfo  {journal} {Phys. Rev. B}\ }\textbf {\bibinfo {volume} {94}},\ \bibinfo {pages} {035135} (\bibinfo {year} {2016})}\BibitemShut {NoStop}%
\bibitem [{\citenamefont {Garc\'{\i}a-Garc\'{\i}a}\ and\ \citenamefont {Verbaarschot}(2016)}]{GarciaGarcia-16}%
  \BibitemOpen
  \bibfield  {author} {\bibinfo {author} {\bibfnamefont {A.~M.}\ \bibnamefont {Garc\'{\i}a-Garc\'{\i}a}}\ and\ \bibinfo {author} {\bibfnamefont {J.~J.~M.}\ \bibnamefont {Verbaarschot}},\ }\bibfield  {title} {\bibinfo {title} {{Spectral and thermodynamic properties of the Sachdev-Ye-Kitaev model}},\ }\href {https://doi.org/10.1103/PhysRevD.94.126010} {\bibfield  {journal} {\bibinfo  {journal} {Phys. Rev. D}\ }\textbf {\bibinfo {volume} {94}},\ \bibinfo {pages} {126010} (\bibinfo {year} {2016})}\BibitemShut {NoStop}%
\bibitem [{\citenamefont {Cotler}\ \emph {et~al.}(2017)\citenamefont {Cotler}, \citenamefont {Gur-Ari}, \citenamefont {Hanada}, \citenamefont {Polchinski}, \citenamefont {Saad}, \citenamefont {Shenker}, \citenamefont {Stanford}, \citenamefont {Streicher},\ and\ \citenamefont {Tezuka}}]{Cotler-17}%
  \BibitemOpen
  \bibfield  {author} {\bibinfo {author} {\bibfnamefont {J.~S.}\ \bibnamefont {Cotler}}, \bibinfo {author} {\bibfnamefont {G.}~\bibnamefont {Gur-Ari}}, \bibinfo {author} {\bibfnamefont {M.}~\bibnamefont {Hanada}}, \bibinfo {author} {\bibfnamefont {J.}~\bibnamefont {Polchinski}}, \bibinfo {author} {\bibfnamefont {P.}~\bibnamefont {Saad}}, \bibinfo {author} {\bibfnamefont {S.~H.}\ \bibnamefont {Shenker}}, \bibinfo {author} {\bibfnamefont {D.}~\bibnamefont {Stanford}}, \bibinfo {author} {\bibfnamefont {A.}~\bibnamefont {Streicher}},\ and\ \bibinfo {author} {\bibfnamefont {M.}~\bibnamefont {Tezuka}},\ }\bibfield  {title} {\bibinfo {title} {{Black holes and random matrices}},\ }\href {https://doi.org/10.1007/JHEP05(2017)118} {\bibfield  {journal} {\bibinfo  {journal} {J. High Energ. Phys.}\ }\textbf {\bibinfo {volume} {2017}}\bibinfo  {number} { (5)},\ \bibinfo {pages} {118}}\BibitemShut {NoStop}%
\bibitem [{\citenamefont {Li}\ \emph {et~al.}(2017)\citenamefont {Li}, \citenamefont {Liu}, \citenamefont {Xin},\ and\ \citenamefont {Zhou}}]{Li-17}%
  \BibitemOpen
\bibfield  {number} {  }\bibfield  {author} {\bibinfo {author} {\bibfnamefont {T.}~\bibnamefont {Li}}, \bibinfo {author} {\bibfnamefont {J.}~\bibnamefont {Liu}}, \bibinfo {author} {\bibfnamefont {Y.}~\bibnamefont {Xin}},\ and\ \bibinfo {author} {\bibfnamefont {Y.}~\bibnamefont {Zhou}},\ }\bibfield  {title} {\bibinfo {title} {{Supersymmetric SYK model and random matrix theory}},\ }\href {https://doi.org/10.1007/JHEP06(2017)111} {\bibfield  {journal} {\bibinfo  {journal} {J. High Energ. Phys.}\ }\textbf {\bibinfo {volume} {2017}}\bibinfo  {number} { (6)},\ \bibinfo {pages} {111}}\BibitemShut {NoStop}%
\bibitem [{\citenamefont {Kanazawa}\ and\ \citenamefont {Wettig}(2017)}]{Kanazawa-17}%
  \BibitemOpen
\bibfield  {number} {  }\bibfield  {author} {\bibinfo {author} {\bibfnamefont {T.}~\bibnamefont {Kanazawa}}\ and\ \bibinfo {author} {\bibfnamefont {T.}~\bibnamefont {Wettig}},\ }\bibfield  {title} {\bibinfo {title} {{Complete random matrix classification of SYK models with $\mathcal{N} = 0, 1$ and $2$ supersymmetry}},\ }\href {https://doi.org/10.1007/JHEP09(2017)050} {\bibfield  {journal} {\bibinfo  {journal} {J. High Energ. Phys.}\ }\textbf {\bibinfo {volume} {2017}}\bibinfo  {number} { (9)},\ \bibinfo {pages} {50}}\BibitemShut {NoStop}%
\bibitem [{\citenamefont {Behrends}\ \emph {et~al.}(2019)\citenamefont {Behrends}, \citenamefont {Bardarson},\ and\ \citenamefont {B\'eri}}]{Behrends-19}%
  \BibitemOpen
\bibfield  {number} {  }\bibfield  {author} {\bibinfo {author} {\bibfnamefont {J.}~\bibnamefont {Behrends}}, \bibinfo {author} {\bibfnamefont {J.~H.}\ \bibnamefont {Bardarson}},\ and\ \bibinfo {author} {\bibfnamefont {B.}~\bibnamefont {B\'eri}},\ }\bibfield  {title} {\bibinfo {title} {{Tenfold way and many-body zero modes in the Sachdev-Ye-Kitaev model}},\ }\href {https://doi.org/10.1103/PhysRevB.99.195123} {\bibfield  {journal} {\bibinfo  {journal} {Phys. Rev. B}\ }\textbf {\bibinfo {volume} {99}},\ \bibinfo {pages} {195123} (\bibinfo {year} {2019})}\BibitemShut {NoStop}%
\bibitem [{\citenamefont {Sun}\ and\ \citenamefont {Ye}(2020)}]{Sun-20}%
  \BibitemOpen
  \bibfield  {author} {\bibinfo {author} {\bibfnamefont {F.}~\bibnamefont {Sun}}\ and\ \bibinfo {author} {\bibfnamefont {J.}~\bibnamefont {Ye}},\ }\bibfield  {title} {\bibinfo {title} {{Periodic Table of the Ordinary and Supersymmetric Sachdev-Ye-Kitaev Models}},\ }\href {https://doi.org/10.1103/PhysRevLett.124.244101} {\bibfield  {journal} {\bibinfo  {journal} {Phys. Rev. Lett.}\ }\textbf {\bibinfo {volume} {124}},\ \bibinfo {pages} {244101} (\bibinfo {year} {2020})}\BibitemShut {NoStop}%
\bibitem [{\citenamefont {Lian}(2023)}]{Lian-23}%
  \BibitemOpen
  \bibfield  {author} {\bibinfo {author} {\bibfnamefont {B.}~\bibnamefont {Lian}},\ }\bibfield  {title} {\bibinfo {title} {{Quantum breakdown model: From many-body localization to chaos with scars}},\ }\href {https://doi.org/10.1103/PhysRevB.107.115171} {\bibfield  {journal} {\bibinfo  {journal} {Phys. Rev. B}\ }\textbf {\bibinfo {volume} {107}},\ \bibinfo {pages} {115171} (\bibinfo {year} {2023})}\BibitemShut {NoStop}%
\bibitem [{\citenamefont {Guan}\ and\ \citenamefont {Katsura}(2026)}]{Guan-Katsura-26}%
  \BibitemOpen
  \bibfield  {author} {\bibinfo {author} {\bibfnamefont {K.}~\bibnamefont {Guan}}\ and\ \bibinfo {author} {\bibfnamefont {H.}~\bibnamefont {Katsura}},\ }\bibfield  {title} {\bibinfo {title} {{Exactly Solvable Disorder-free Quantum Breakdown Model: Spectrum, Thermodynamics, and Dynamics}},\ }\Eprint {https://arxiv.org/abs/arXiv:2603.17379} {arXiv:2603.17379}  (\bibinfo {year} {2026})\BibitemShut {NoStop}%
\bibitem [{\citenamefont {Jonay}\ \emph {et~al.}(2026)\citenamefont {Jonay}, \citenamefont {Kim}, \citenamefont {Pollmann},\ and\ \citenamefont {Altland}}]{Jonay-26}%
  \BibitemOpen
  \bibfield  {author} {\bibinfo {author} {\bibfnamefont {C.}~\bibnamefont {Jonay}}, \bibinfo {author} {\bibfnamefont {K.~W.}\ \bibnamefont {Kim}}, \bibinfo {author} {\bibfnamefont {F.}~\bibnamefont {Pollmann}},\ and\ \bibinfo {author} {\bibfnamefont {A.}~\bibnamefont {Altland}},\ }\bibfield  {title} {\bibinfo {title} {{Quantum Chaos with a Macroscopic Zero-Mode Sector}},\ }\Eprint {https://arxiv.org/abs/arXiv:2607.09504} {arXiv:2607.09504}  (\bibinfo {year} {2026})\BibitemShut {NoStop}%
\bibitem [{\citenamefont {Turner}\ \emph {et~al.}(2018)\citenamefont {Turner}, \citenamefont {Michailidis}, \citenamefont {Abanin}, \citenamefont {Serbyn},\ and\ \citenamefont {Papi\ifmmode~\acute{c}\else \'{c}\fi{}}}]{Turner-18}%
  \BibitemOpen
  \bibfield  {author} {\bibinfo {author} {\bibfnamefont {C.~J.}\ \bibnamefont {Turner}}, \bibinfo {author} {\bibfnamefont {A.~A.}\ \bibnamefont {Michailidis}}, \bibinfo {author} {\bibfnamefont {D.~A.}\ \bibnamefont {Abanin}}, \bibinfo {author} {\bibfnamefont {M.}~\bibnamefont {Serbyn}},\ and\ \bibinfo {author} {\bibfnamefont {Z.}~\bibnamefont {Papi\ifmmode~\acute{c}\else \'{c}\fi{}}},\ }\bibfield  {title} {\bibinfo {title} {{Quantum scarred eigenstates in a Rydberg atom chain: Entanglement, breakdown of thermalization, and stability to perturbations}},\ }\href {https://doi.org/10.1103/PhysRevB.98.155134} {\bibfield  {journal} {\bibinfo  {journal} {Phys. Rev. B}\ }\textbf {\bibinfo {volume} {98}},\ \bibinfo {pages} {155134} (\bibinfo {year} {2018})}\BibitemShut {NoStop}%
\bibitem [{\citenamefont {Banerjee}\ and\ \citenamefont {Sen}(2021)}]{Banerjee-21}%
  \BibitemOpen
  \bibfield  {author} {\bibinfo {author} {\bibfnamefont {D.}~\bibnamefont {Banerjee}}\ and\ \bibinfo {author} {\bibfnamefont {A.}~\bibnamefont {Sen}},\ }\bibfield  {title} {\bibinfo {title} {{Quantum Scars from Zero Modes in an Abelian Lattice Gauge Theory on Ladders}},\ }\href {https://doi.org/10.1103/PhysRevLett.126.220601} {\bibfield  {journal} {\bibinfo  {journal} {Phys. Rev. Lett.}\ }\textbf {\bibinfo {volume} {126}},\ \bibinfo {pages} {220601} (\bibinfo {year} {2021})}\BibitemShut {NoStop}%
\bibitem [{\citenamefont {Buijsman}(2022)}]{Buijsman-22}%
  \BibitemOpen
  \bibfield  {author} {\bibinfo {author} {\bibfnamefont {W.}~\bibnamefont {Buijsman}},\ }\bibfield  {title} {\bibinfo {title} {{Number of zero-energy eigenstates in the PXP model}},\ }\href {https://doi.org/10.1103/PhysRevB.106.045104} {\bibfield  {journal} {\bibinfo  {journal} {Phys. Rev. B}\ }\textbf {\bibinfo {volume} {106}},\ \bibinfo {pages} {045104} (\bibinfo {year} {2022})}\BibitemShut {NoStop}%
\bibitem [{\citenamefont {Sannomiya}\ \emph {et~al.}(2017)\citenamefont {Sannomiya}, \citenamefont {Katsura},\ and\ \citenamefont {Nakayama}}]{Sannomiya-17}%
  \BibitemOpen
  \bibfield  {author} {\bibinfo {author} {\bibfnamefont {N.}~\bibnamefont {Sannomiya}}, \bibinfo {author} {\bibfnamefont {H.}~\bibnamefont {Katsura}},\ and\ \bibinfo {author} {\bibfnamefont {Y.}~\bibnamefont {Nakayama}},\ }\bibfield  {title} {\bibinfo {title} {{Supersymmetry breaking and Nambu-Goldstone fermions with cubic dispersion}},\ }\href {https://doi.org/10.1103/PhysRevD.95.065001} {\bibfield  {journal} {\bibinfo  {journal} {Phys. Rev. D}\ }\textbf {\bibinfo {volume} {95}},\ \bibinfo {pages} {065001} (\bibinfo {year} {2017})}\BibitemShut {NoStop}%
\bibitem [{\citenamefont {Schecter}\ and\ \citenamefont {Iadecola}(2018)}]{Schecter-18}%
  \BibitemOpen
  \bibfield  {author} {\bibinfo {author} {\bibfnamefont {M.}~\bibnamefont {Schecter}}\ and\ \bibinfo {author} {\bibfnamefont {T.}~\bibnamefont {Iadecola}},\ }\bibfield  {title} {\bibinfo {title} {{Many-body spectral reflection symmetry and protected infinite-temperature degeneracy}},\ }\href {https://doi.org/10.1103/PhysRevB.98.035139} {\bibfield  {journal} {\bibinfo  {journal} {Phys. Rev. B}\ }\textbf {\bibinfo {volume} {98}},\ \bibinfo {pages} {035139} (\bibinfo {year} {2018})}\BibitemShut {NoStop}%
\bibitem [{\citenamefont {Iyoda}\ \emph {et~al.}(2018)\citenamefont {Iyoda}, \citenamefont {Katsura},\ and\ \citenamefont {Sagawa}}]{Iyoda-18}%
  \BibitemOpen
  \bibfield  {author} {\bibinfo {author} {\bibfnamefont {E.}~\bibnamefont {Iyoda}}, \bibinfo {author} {\bibfnamefont {H.}~\bibnamefont {Katsura}},\ and\ \bibinfo {author} {\bibfnamefont {T.}~\bibnamefont {Sagawa}},\ }\bibfield  {title} {\bibinfo {title} {{Effective dimension, level statistics, and integrability of Sachdev-Ye-Kitaev-like models}},\ }\href {https://doi.org/10.1103/PhysRevD.98.086020} {\bibfield  {journal} {\bibinfo  {journal} {Phys. Rev. D}\ }\textbf {\bibinfo {volume} {98}},\ \bibinfo {pages} {086020} (\bibinfo {year} {2018})}\BibitemShut {NoStop}%
\bibitem [{\citenamefont {Oganesyan}\ and\ \citenamefont {Huse}(2007)}]{Oganesyan-07}%
  \BibitemOpen
  \bibfield  {author} {\bibinfo {author} {\bibfnamefont {V.}~\bibnamefont {Oganesyan}}\ and\ \bibinfo {author} {\bibfnamefont {D.~A.}\ \bibnamefont {Huse}},\ }\bibfield  {title} {\bibinfo {title} {{Localization of interacting fermions at high temperature}},\ }\href {https://doi.org/10.1103/PhysRevB.75.155111} {\bibfield  {journal} {\bibinfo  {journal} {Phys. Rev. B}\ }\textbf {\bibinfo {volume} {75}},\ \bibinfo {pages} {155111} (\bibinfo {year} {2007})}\BibitemShut {NoStop}%
\bibitem [{\citenamefont {Atas}\ \emph {et~al.}(2013)\citenamefont {Atas}, \citenamefont {Bogomolny}, \citenamefont {Giraud},\ and\ \citenamefont {Roux}}]{Atas-13}%
  \BibitemOpen
  \bibfield  {author} {\bibinfo {author} {\bibfnamefont {Y.~Y.}\ \bibnamefont {Atas}}, \bibinfo {author} {\bibfnamefont {E.}~\bibnamefont {Bogomolny}}, \bibinfo {author} {\bibfnamefont {O.}~\bibnamefont {Giraud}},\ and\ \bibinfo {author} {\bibfnamefont {G.}~\bibnamefont {Roux}},\ }\bibfield  {title} {\bibinfo {title} {{Distribution of the Ratio of Consecutive Level Spacings in Random Matrix Ensembles}},\ }\href {https://doi.org/10.1103/PhysRevLett.110.084101} {\bibfield  {journal} {\bibinfo  {journal} {Phys. Rev. Lett.}\ }\textbf {\bibinfo {volume} {110}},\ \bibinfo {pages} {084101} (\bibinfo {year} {2013})}\BibitemShut {NoStop}%
\bibitem [{\citenamefont {Nishigaki}\ \emph {et~al.}(1998)\citenamefont {Nishigaki}, \citenamefont {Damgaard},\ and\ \citenamefont {Wettig}}]{Nishigaki-98}%
  \BibitemOpen
  \bibfield  {author} {\bibinfo {author} {\bibfnamefont {S.~M.}\ \bibnamefont {Nishigaki}}, \bibinfo {author} {\bibfnamefont {P.~H.}\ \bibnamefont {Damgaard}},\ and\ \bibinfo {author} {\bibfnamefont {T.}~\bibnamefont {Wettig}},\ }\bibfield  {title} {\bibinfo {title} {{Smallest Dirac eigenvalue distribution from random matrix theory}},\ }\href {https://doi.org/10.1103/PhysRevD.58.087704} {\bibfield  {journal} {\bibinfo  {journal} {Phys. Rev. D}\ }\textbf {\bibinfo {volume} {58}},\ \bibinfo {pages} {087704} (\bibinfo {year} {1998})}\BibitemShut {NoStop}%
\bibitem [{\citenamefont {Damgaard}\ and\ \citenamefont {Nishigaki}(2001)}]{Damgaard-01}%
  \BibitemOpen
  \bibfield  {author} {\bibinfo {author} {\bibfnamefont {P.~H.}\ \bibnamefont {Damgaard}}\ and\ \bibinfo {author} {\bibfnamefont {S.~M.}\ \bibnamefont {Nishigaki}},\ }\bibfield  {title} {\bibinfo {title} {{Distribution of the $k$th smallest Dirac operator eigenvalue}},\ }\href {https://doi.org/10.1103/PhysRevD.63.045012} {\bibfield  {journal} {\bibinfo  {journal} {Phys. Rev. D}\ }\textbf {\bibinfo {volume} {63}},\ \bibinfo {pages} {045012} (\bibinfo {year} {2001})}\BibitemShut {NoStop}%
\bibitem [{\citenamefont {Nagao}\ and\ \citenamefont {Forrester}(1995)}]{Nagao-95}%
  \BibitemOpen
  \bibfield  {author} {\bibinfo {author} {\bibfnamefont {T.}~\bibnamefont {Nagao}}\ and\ \bibinfo {author} {\bibfnamefont {P.~J.}\ \bibnamefont {Forrester}},\ }\bibfield  {title} {\bibinfo {title} {{Asymptotic correlations at the spectrum edge of random matrices}},\ }\href {https://doi.org/https://doi.org/10.1016/0550-3213(94)00545-P} {\bibfield  {journal} {\bibinfo  {journal} {Nucl. Phys. B}\ }\textbf {\bibinfo {volume} {435}},\ \bibinfo {pages} {401} (\bibinfo {year} {1995})}\BibitemShut {NoStop}%
\bibitem [{\citenamefont {Nagao}\ and\ \citenamefont {Nishigaki}(2000)}]{Nagao-00}%
  \BibitemOpen
  \bibfield  {author} {\bibinfo {author} {\bibfnamefont {T.}~\bibnamefont {Nagao}}\ and\ \bibinfo {author} {\bibfnamefont {S.~M.}\ \bibnamefont {Nishigaki}},\ }\bibfield  {title} {\bibinfo {title} {{Massive chiral random matrix ensembles at $\beta=1$ and $4$: QCD Dirac operator spectra}},\ }\href {https://doi.org/10.1103/PhysRevD.62.065007} {\bibfield  {journal} {\bibinfo  {journal} {Phys. Rev. D}\ }\textbf {\bibinfo {volume} {62}},\ \bibinfo {pages} {065007} (\bibinfo {year} {2000})}\BibitemShut {NoStop}%
\bibitem [{\citenamefont {Nagao}\ and\ \citenamefont {Forrester}(1998)}]{Nagao-98}%
  \BibitemOpen
  \bibfield  {author} {\bibinfo {author} {\bibfnamefont {T.}~\bibnamefont {Nagao}}\ and\ \bibinfo {author} {\bibfnamefont {P.~J.}\ \bibnamefont {Forrester}},\ }\bibfield  {title} {\bibinfo {title} {{The smallest eigenvalue distribution at the spectrum edge of random matrices}},\ }\href {https://doi.org/https://doi.org/10.1016/S0550-3213(97)00670-6} {\bibfield  {journal} {\bibinfo  {journal} {Nucl. Phys. B}\ }\textbf {\bibinfo {volume} {509}},\ \bibinfo {pages} {561} (\bibinfo {year} {1998})}\BibitemShut {NoStop}%
\bibitem [{\citenamefont {Ivanov}(2002)}]{Ivanov-02}%
  \BibitemOpen
  \bibfield  {author} {\bibinfo {author} {\bibfnamefont {D.~A.}\ \bibnamefont {Ivanov}},\ }\bibfield  {title} {\bibinfo {title} {{The supersymmetric technique for random-matrix ensembles with zero eigenvalues}},\ }\href {https://doi.org/10.1063/1.1423765} {\bibfield  {journal} {\bibinfo  {journal} {J. Math. Phys.}\ }\textbf {\bibinfo {volume} {43}},\ \bibinfo {pages} {126} (\bibinfo {year} {2002})}\BibitemShut {NoStop}%
\bibitem [{\citenamefont {{V. A. Mar\v{c}enko and L. A. Pastur}}(1967)}]{Marchenko-Pastur-67}%
  \BibitemOpen
  \bibfield  {author} {\bibinfo {author} {\bibnamefont {{V. A. Mar\v{c}enko and L. A. Pastur}}},\ }\bibfield  {title} {\bibinfo {title} {{Distribution of eigenvalues for some sets of random matrices}},\ }\href {https://doi.org/10.1070/SM1967v001n04ABEH001994} {\bibfield  {journal} {\bibinfo  {journal} {Math. USSR-Sbornik}\ }\textbf {\bibinfo {volume} {1}},\ \bibinfo {pages} {457} (\bibinfo {year} {1967})}\BibitemShut {NoStop}%
\bibitem [{\citenamefont {Fidkowski}\ and\ \citenamefont {Kitaev}(2010)}]{Fidkowski-Kitaev-10}%
  \BibitemOpen
  \bibfield  {author} {\bibinfo {author} {\bibfnamefont {L.}~\bibnamefont {Fidkowski}}\ and\ \bibinfo {author} {\bibfnamefont {A.}~\bibnamefont {Kitaev}},\ }\bibfield  {title} {\bibinfo {title} {{Effects of interactions on the topological classification of free fermion systems}},\ }\href {https://doi.org/10.1103/PhysRevB.81.134509} {\bibfield  {journal} {\bibinfo  {journal} {Phys. Rev. B}\ }\textbf {\bibinfo {volume} {81}},\ \bibinfo {pages} {134509} (\bibinfo {year} {2010})}\BibitemShut {NoStop}%
\bibitem [{\citenamefont {Fidkowski}\ and\ \citenamefont {Kitaev}(2011)}]{Fidkowski-Kitaev-11}%
  \BibitemOpen
  \bibfield  {author} {\bibinfo {author} {\bibfnamefont {L.}~\bibnamefont {Fidkowski}}\ and\ \bibinfo {author} {\bibfnamefont {A.}~\bibnamefont {Kitaev}},\ }\bibfield  {title} {\bibinfo {title} {{Topological phases of fermions in one dimension}},\ }\href {https://doi.org/10.1103/PhysRevB.83.075103} {\bibfield  {journal} {\bibinfo  {journal} {Phys. Rev. B}\ }\textbf {\bibinfo {volume} {83}},\ \bibinfo {pages} {075103} (\bibinfo {year} {2011})}\BibitemShut {NoStop}%
\bibitem [{\citenamefont {Turner}\ \emph {et~al.}(2011)\citenamefont {Turner}, \citenamefont {Pollmann},\ and\ \citenamefont {Berg}}]{Turner-11}%
  \BibitemOpen
  \bibfield  {author} {\bibinfo {author} {\bibfnamefont {A.~M.}\ \bibnamefont {Turner}}, \bibinfo {author} {\bibfnamefont {F.}~\bibnamefont {Pollmann}},\ and\ \bibinfo {author} {\bibfnamefont {E.}~\bibnamefont {Berg}},\ }\bibfield  {title} {\bibinfo {title} {{Topological phases of one-dimensional fermions: An entanglement point of view}},\ }\href {https://doi.org/10.1103/PhysRevB.83.075102} {\bibfield  {journal} {\bibinfo  {journal} {Phys. Rev. B}\ }\textbf {\bibinfo {volume} {83}},\ \bibinfo {pages} {075102} (\bibinfo {year} {2011})}\BibitemShut {NoStop}%
\bibitem [{\citenamefont {Sanada}\ \emph {et~al.}(2023)\citenamefont {Sanada}, \citenamefont {Miao},\ and\ \citenamefont {Katsura}}]{Sanada-23}%
  \BibitemOpen
  \bibfield  {author} {\bibinfo {author} {\bibfnamefont {K.}~\bibnamefont {Sanada}}, \bibinfo {author} {\bibfnamefont {Y.}~\bibnamefont {Miao}},\ and\ \bibinfo {author} {\bibfnamefont {H.}~\bibnamefont {Katsura}},\ }\bibfield  {title} {\bibinfo {title} {{Quantum many-body scars in spin models with multibody interactions}},\ }\href {https://doi.org/10.1103/PhysRevB.108.155102} {\bibfield  {journal} {\bibinfo  {journal} {Phys. Rev. B}\ }\textbf {\bibinfo {volume} {108}},\ \bibinfo {pages} {155102} (\bibinfo {year} {2023})}\BibitemShut {NoStop}%
\end{thebibliography}%
\let\addcontentsline\oldaddcontentsline

\end{document}